\documentclass[letterpaper,english,reprint,aps,prd,superscriptaddress, longbibliography]{revtex4-2}
\usepackage[T1]{fontenc}
\usepackage[utf8]{inputenc}
\usepackage{color}
\usepackage{babel}
\usepackage{prettyref}
\usepackage{bm}
\usepackage{amsmath}
\usepackage{amssymb}
\usepackage{graphicx}
\usepackage[pdfusetitle,
 bookmarks=true,bookmarksnumbered=true,bookmarksopen=true,bookmarksopenlevel=1,
 breaklinks=true,pdfborder={0 0 0},pdfborderstyle={},backref=false,colorlinks=true]
 {hyperref}
\hypersetup{
 colorlinks=true,linkcolor=blue,citecolor=blue,urlcolor=blue}

\makeatletter

\pdfpageheight\paperheight
\pdfpagewidth\paperwidth

\allowdisplaybreaks

\makeatother

\begin{document}
\preprint{}
\title{Holographic Turbulence and Numerical Estimate of the Fractal Dimension of the Turbulent Horizon}
\author{Jia Du}
\email{dujia22@mails.ucas.ac.cn}

\affiliation{School of Physics, University of Chinese Academy of Sciences, Beijing 100190, China}

\author{Yu Tian}
\email{ytian@ucas.ac.cn}
\affiliation{School of Physics, University of Chinese Academy of Sciences, Beijing 100190, China}
\affiliation{Institute of Theoretical Physics, Chinese Academy of Sciences, Beijing
100190, China }

\author{Hongbao Zhang}
\email{hongbaozhang@bnu.edu.cn}

\affiliation{School of Physics and Astronomy, Beijing Normal University, Beijing
100875, China}
\affiliation{Key Laboratory of Multiscale Spin Physics, Ministry of Education,
Beijing Normal University, Beijing 100875, China}
\date{\today}

\begin{abstract}
We numerically study two-dimensional turbulence driven by a scalar operator within the framework of the AdS/CFT correspondence, where the external driving source is used to sustain a quasi-steady turbulent state. We propose a simple and efficient evolution scheme within the Bondi-Sachs formalism. Applying this scheme to numerically solve the full nonlinear equations of motion, we obtain a turbulent black hole in asymptotically $\mathrm{AdS}_4$ spacetime. The inverse energy cascade and the corresponding energy spectrum of both decaying and driven dual turbulence are analyzed. The scalar driving leads to a compressible-energy-dominated flow, and the corresponding power law scaling, $E(k)\propto k^{-1.79}$, agrees well with previous simulations of two-dimensional turbulence in weakly coupled compressible fluids in fluid dynamics. This differs from the Kolmogorov $-5/3$ scaling law. Furthermore, we perform a direct numerical estimate of the fractal structure of the turbulent black hole, obtaining a fractal dimension $D\approx 2.65$, which suggests an interesting universality
in the fractal dimension.
\end{abstract}
\maketitle

\section{Introduction\label{sec:Introduction}}

Turbulence is a universal, chaotic, and highly complex phenomenon
in nature, playing central roles across a broad range of physical
systems. Despite being investigated extensively over centuries through
both experimental and theoretical methods, e.g. \citep{ExperimentalInvestigationCircumstances1883,LocalStructureTurbulence1941,DistributionEnergySpectrum1941,DissipationEnergyLocally1941,RefinementPreviousHypotheses1962,InertialRangesTwoDimensional1967,FluidMechanics1987,TurbulenceLegacyKolmogorov2009,TwoDimensionalTurbulence2012},
it remains an important and unresolved problem in modern physics.

Recent advances in black hole physics provide a novel perspective
on this longstanding challenge. It has been remarkably shown that
in the long wavelength limit the relativistic fluid dynamics in the
$d$-dimensional conformal field theory (CFT) can correspond
to the black hole dynamics in the $\left(d+1\right)$-dimensional asymptotically
anti-de Sitter (AdS) spacetime \citep{AdSCFTCorrespondence2002a,NonlinearFluidDynamics2008,RelativisticViscousHydrodynamics2008,GravityHydrodynamicsLectures2009,FluidGravityCorrespondence2011,FluidGravityCorrespondence2011a},
within the framework of the AdS/CFT correspondence \citep{LargeLimitSuperconformal1999,GaugeTheoryCorrelators1998,SitterSpaceHolography1998}.
This fluid/gravity duality connects two important dynamical systems
and raises an interesting question: \emph{what can we learn about
turbulent fluid dynamics from gravity and vice versa?}

Indeed, much interesting progress has been made, including relations
between the Einstein equation and the Navier-Stokes equation \citep{BlackHoleDynamics2008a,LargeRotatingAds2008,IncompressibleNonRelativisticNavierStokes2009,IncompressibleNavierStokesEquations2009,RelativisticCFTHydrodynamics2010,GravityGeometrizationTurbulence2010,BlackHolesIncompressible2012},
relativistic fluid dynamics and black hole dynamics \citep{ForcedFluidDynamics2009,TurbulentFlowsRelativistic2012,HolographicForcedFluid2012,ForcedFluidDynamics2014,HolographicPathTurbulent2014,DynamicalSpacetimesNumerical2014,ScalingRelationsTwoDimensional2015,NumericalMeasurementsScaling2017,HolographicTurbulenceLarge2018,DrivenBlackHoles2021,TwodimensionalFluidsTheir2019,StochasticGravityTurbulence2021,HolographicTurbulenceRandom2024},
the fluid entropy current and the horizon area increase theorem
\citep{LocalFluidDynamical2008,RelativisticCFTHydrodynamics2010},
macroscopic views of hydrodynamics provided by CFT \citep{ConformalFieldTheory2008}
and so on. Even beyond the fluid/gravity duality, some turbulent behaviors
are also found in gravity. For instance, chaos around AdS black holes \citep{ChaosGaugeGravity2010}, energy cascade in unstable AdS spacetimes \citep{WeaklyTurbulentInstability2011,KolmogorovZakharovSpectrumAdS2013}, parametric
resonant turbulent black holes \citep{TurbulentBlackHoles2015} and
turbulent modes in gravitational waves \citep{TurbulenceWeakGravitational2017,DirectEvidenceDual2021,TurbulenceSpacetimesStable2024,WeaklyTurbulentSaturation2025,AsymmetricDualCascade2025,EmergentTurbulenceNonlinear2025,TheoryGravitationalWave2025}.

In the pioneering work of \citep{HolographicTurbulence2014}, it was realized that two-dimensional turbulent flows can emerge directly
from the nonlinear dynamics of turbulent black holes. They showed that an \emph{inverse energy cascade} exists, where the energy of the turbulent flow is transferred from short to long wavelengths. They claimed that the energy spectrum of their turbulence scaled as $k^{-5/3}$, captured by the well-known theory of Kolmogorov \citep{LocalStructureTurbulence1941,DissipationEnergyLocally1941}.
Based on this Kolmogorov's scaling, a geometric fractal structure with dimension $D=d+1/3$ was proposed for the horizon of $d$-dimensional turbulent black holes in a steady state.

However, both their obtained scaling law and the proposed fractal dimension are not fully justified. First, Kolmogorov's theory derived from dimensional analysis
is restricted to incompressible and non-relativistic steady-state fluids, while the flow considered in \citep{HolographicTurbulence2014} should be compressible and starts in a relativistic regime. Also, in the absence of an external driving force, their turbulent flow freely decays, leading to a short inertial range and transient energy spectrum, as we will demonstrate. This limited inertial range covers few scales and is possibly insufficient to make a reliable measurement of the fractal dimension. Second, their fractal dimension exceeds the topological dimension of the black horizon and may not be valid \citep{FractalDimensionTurbulent2017}.
Therefore, one may naturally wonder whether a better estimate of the fractal dimension is possible, and whether the turbulent scaling behavior would exist or differ in a steady state.

Although several studies have tried to address the above questions, they all have some limitations. In \citep{FractalDimensionTurbulent2017}, they considered the boundary ideal conformal fluid and constructed the bulk metric only up to the ideal order, from which their fractal dimensions were calculated. But as an ideal conformal fluid does not dissipate, the dissipation in their fluid evolution equations was introduced by numerical considerations \citep{NumericalMeasurementsScaling2017}. Such a dissipation term is neither derived from the fluid/gravity duality, nor from the relativistic hydrodynamics \citep{TurbulentFlowsRelativistic2012,HolographicPathTurbulent2014,DynamicalSpacetimesNumerical2014}. Furthermore, since turbulence involves energy cascade across different scales, it is significant to go beyond a derivative expansion, as was first performed by full nonlinear evolution in \citep{HolographicTurbulence2014}. However, the turbulence observed in \citep{HolographicTurbulence2014} is freely decaying with transient scaling and short inertial range. To rescue these deficiencies, \citep{StochasticGravityTurbulence2021,HolographicTurbulenceRandom2024} produce steady turbulence by stochastically driving the boundary metric; their numerical resolution, however, does not suffice to measure the fractal dimension and the inertial range is too small for one to extract accurate results. Therefore, it is necessary to have a full nonlinear bulk evolution which goes beyond the derivative expansion and have a sufficiently wide inertial range to estimate the fractal dimension.

In this work, with the above motivations, we extend the full nonlinear
investigations by numerically constructing a $\left(3+1\right)$-dimensional turbulent black hole which is dual to a $\left(2+1\right)$-dimensional compressible fluid. A massive scalar field is introduced and its boundary source is constructed by periodic interpolation between randomly drawn values, making it a deterministic external driving force, similar to \citep{DrivenBlackHoles2021}. An efficient evolution scheme is employed in the Bondi-Sachs formalism specialized for AdS spacetime, using $512$ Fourier modes per spatial direction, yielding a steady inertial range $k\in[10,65]$. This is the widest attained so far in full nonlinear holographic turbulence, compared to $k\in[3,14]$ in \cite{StochasticGravityTurbulence2021,HolographicTurbulenceRandom2024} and $k\in[5,10]$ in \cite{HolographicTurbulence2014}. This facilitates the estimation of the fractal dimension of the black hole horizon. After statistical averages, we obtain the fractal dimension of the turbulent black hole using the methodology introduced in \citep{FractalDimensionTurbulent2017} and its corresponding energy spectrum scaling as

\begin{equation}
D\approx2.65\pm0.02,\quad E\left(k\right)\sim k^{-1.79\pm0.03}.
\end{equation}
 Comparing this value of the fractal dimension with that reported in \citep{FractalDimensionTurbulent2017}, we observe a surprising universality in $D$ despite the different scaling behavior of the energy spectrum. Although this is a very interesting result, we will not investigate the origin of this universality in the present work. We find that the energy spectrum of the driven turbulence is close but not exactly equal to Kolmogorov's $k^{-5/3}$ scaling. After a Helmholtz decomposition which divides the energy spectrum into the compressible and incompressible components, we find that the compressible component dominates the turbulent energy spectrum with the two components scaling as
\begin{equation}
E_{\mathrm{c}}\sim k^{-1.80\pm0.03},\quad E_{\mathrm{i}}\left(k\right)\sim k^{-1.99\pm0.03},
\end{equation}
respectively. These two scaling powers are consistent with the simulations of two-dimensional compressible fluids \citep{EnergyTransferSpectra2019}.

The rest of this paper is structured as follows. In \prettyref{sec:TheDualGravModel}, we introduce our gravitational model and the hierarchical structure of the Einstein equation. Technical details are deferred to Appendix \ref{sec:appendix_Holographic-Renormalization},\ref{sec:appendix_numerical_details}. In \prettyref{sec:DecayingTurb}, we investigate the decaying turbulence dual to the vacuum Einstein gravity. In \prettyref{sec:ForcedTurb}, we randomly drive the fluid into a quasi-steady turbulent state and analyze the kinetic spectrum. In \prettyref{sec:fractal_dim_turbulent_BH}, we estimate the fractal dimension of the turbulent black holes. In \prettyref{sec:SumDiscussion}, we conclude our paper with a summary and some discussions. Throughout the paper, the speed of light $c$ and gravitational constant $G$ are set to unity. The wavenumber is expressed in units of $2\pi/L$ where $L$ denotes the periodicity of the boundary spatial directions.

\section{The Dual Gravitational Model \label{sec:TheDualGravModel}}

\subsection{The Gravitational Model}

We consider Einstein gravity minimally coupled to a real scalar
field in the four-dimensional asymptotically anti-de Sitter spacetime
($\mathrm{AdS_{4}}$), described by the action
\begin{equation}
S=\frac{1}{2\kappa^{2}}\int d^{4}x\sqrt{-g}\bigg[R-2\Lambda-\frac{1}{2}\left(\nabla\phi\right)^{2}-V\left(\phi\right)\bigg],\label{eq:einstein_scalar_action}
\end{equation}
where the negative cosmological constant is $\Lambda=-3/\ell^{2}$,
the gravitational constant is $\kappa^{2}=8\pi G_{4}$ and the asymptotic
AdS radius is $\ell$. Then the Einstein equations and the Klein-Gordon equation for the scalar field can be obtained by varying \prettyref{eq:einstein_scalar_action},
\begin{align}
R_{\mu\nu}-\frac{1}{2}g_{\mu\nu}\left(R-2\Lambda\right) & =T_{\mu\nu}\left(\phi\right),\label{eq:EFEs}\\
\nabla_{\mu}\nabla^{\mu}\phi & =\frac{dV\left(\phi\right)}{d\phi},\label{eq:KG}
\end{align}
with the energy-momentum tensor given by 
\begin{equation}
T_{\mu\nu}=\frac{1}{2}\left[\nabla_{\mu}\phi\nabla_{\nu}\phi-g_{\mu\nu}\left(\frac{1}{2}\nabla_{\mu}\phi\nabla^{\mu}\phi+V\left(\phi\right)\right)\right].
\end{equation}
For simplicity, in this paper, we choose a massive scalar field with
no interacting terms with the potential given by
\begin{equation}
V(\phi)=\frac{1}{2}m^{2}\phi^{2},\quad m^{2}=-\frac{2}{\ell^{2}}.
\end{equation}
The mass of the scalar is chosen such that it satisfies the Breitenlohner-Freedman
bound \citep{StabilityGaugedExtended1982}, which leads to stable
black hole solutions in $\mathrm{AdS}_{4}$ spacetime. Moreover, this
choice avoids the logarithmic terms which are singular at the conformal boundary \citep{HolographicReconstructionSpacetime2001,ThermodynamicsAsymptoticallyLocally2005} in the asymptotic structure of
\prettyref{eq:EFEs} and \prettyref{eq:KG} .

\subsection{Equations of Motion in Bondi-Sachs Formalism\label{sub:eom_BS}}

\begin{figure}
\centering
\includegraphics[scale=0.5]{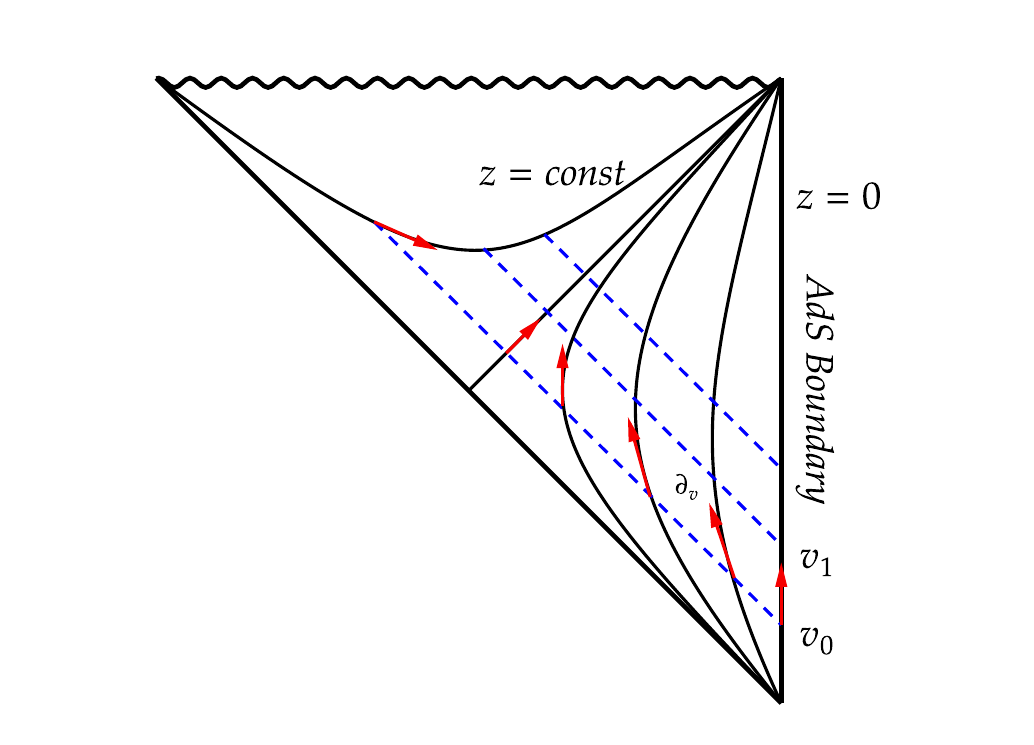}
\caption{Schematic foliation of $\mathrm{AdS_{4}}$ spacetime. Initial data
are given on the ingoing null hypersurfaces (blue dashed lines) at
$v_{0}$ and radial domain is chosen as hypersurfaces between $z=\mathrm{const}$
and the AdS boundary $z=0$. The evolution repeatedly follows along
the vector $\partial_{v}$ (red arrows) from one null hypersurface
into the next.\label{fig:schematic_evolution_fig}}
\end{figure}

The spacetime is foliated by null hypersurfaces as schematically shown in Figure \ref{fig:schematic_evolution_fig}. Specifically, we choose to work in the Bondi-Sachs gauge which exhibits several advantages elaborated in Appendix \ref{sec:appendix_numerical_details}.
The most general metric form in this gauge can be written as 
\begin{multline}
ds^{2}=\frac{\ell^{2}}{z^{2}}\big[-fe^{-\chi}dv^{2}-2e^{-\chi}dvdz\\
+h_{ij}\left(dx^{i}-\xi^{i}dv\right)\left(dx^{j}-\xi^{j}dv\right)\big],\label{eq:Bondi-Sachs-General-Metric}
\end{multline}
where all metric fields are functions of $\left(v,z,x^{i}\right)$
and the components of the shift vector $\xi^{i}$ are denoted as $\left(\xi,\eta\right)$.
The conformal boundary is compactified from $r=\infty$ to $z=0$
through the mapping $z:=\ell^{2}/r$. For the physics we are interested in this work, we consider $\det h_{ij}=1$ in \prettyref{eq:Bondi-Sachs-General-Metric} and parameterize the spatial metric as
\begin{equation}
h_{ij}=\left(\begin{array}{cc}
e^{B}\cosh C & \sinh C\\
\sinh C & e^{-B}\cosh C
\end{array}\right).
\label{eq:spatial_metric}
\end{equation}
Then, the Einstein equations
\begin{equation}
\partial_{z}\chi=S_{\chi}\left[h_{ij},\phi\right]\label{eq:eom_chi},
\end{equation}
\begin{equation}
z^{2}\partial_{z}P_{i}=S_{P_{i}}\left[\chi,h_{ij},\phi\right],\label{eq:eom_Pi}
\end{equation}
\begin{equation}
\left(z\partial_{z}-3\right)f=S_{f}\left[\chi,h_{ij},\phi,P_{i},\xi^{i}\right],\label{eq:eom_f}
\end{equation}
\begin{multline}
\left(z\partial_{z}-1\right)\partial_{v}h_{ij}+\frac{z}{2}\partial_{z}\left(h_{ik}h_{jl}\right)\partial_{v}h^{kl}\\
=S_{h_{ij}}\left[\chi,h_{ij},\phi,P_{i},\xi^{i},f\right],\label{eq:eom_hij}
\end{multline}
and the Klein-Gordon equation 
\begin{multline}
\left(z\partial_{z}-1\right)\partial_{v}\phi=\frac{1}{2}z^{3}\partial_{z}\left(\frac{f\partial_{z}\phi-\xi^{i}\partial_{i}\phi}{z^{2}}\right)\\
+\frac{1}{2}z\partial_{i}(-\xi^{i}\phi+\Theta^{ij}\partial_{j}\phi)-\frac{1}{2}\frac{e^{-\chi}}{z}\frac{\partial V\left(\phi\right)}{\partial\phi},\label{eq:KG-Equation}
\end{multline}
consist of the equations of motion
for our Einstein-scalar system and manifestly show a nested structure.
For convenience, we have defined
\begin{eqnarray}
P_{i} & := & \frac{1}{2z^{2}}\Theta_{ij}\partial_{z}\xi^{j},\quad\Theta_{ij}:=e^{\chi}h_{ij}.\label{eq:eom_for_xi}
\end{eqnarray}
The right-hand terms exhibit their dependence on the metric fields and their derivatives which can be straightforwardly obtained. Through holographic renormalization, the dual fluid equations of motion can be easily obtained as
\begin{equation}
\nabla_{a}\left\langle T_{\;\;b}^{a}\right\rangle =\left\langle O_{\phi}\right\rangle \nabla_{b}\phi_{1},\quad\left\langle O_{\phi}\right\rangle :=\frac{1}{2}\left(\phi_{2}-\partial_{v}\phi_{1}\right).\label{eq:bdry_fluid_EOM}
\end{equation}
The details of the evolution scheme, including our decoupling procedure for equations \eqref{eq:eom_hij}, the chosen integration order, and the construction of the near-boundary series solution, are detailed in Appendix \ref{sec:appendix_numerical_details}.

\section{Decaying Turbulence\label{sec:DecayingTurb}}

\begin{figure*}
\centering
\includegraphics[scale=0.31]{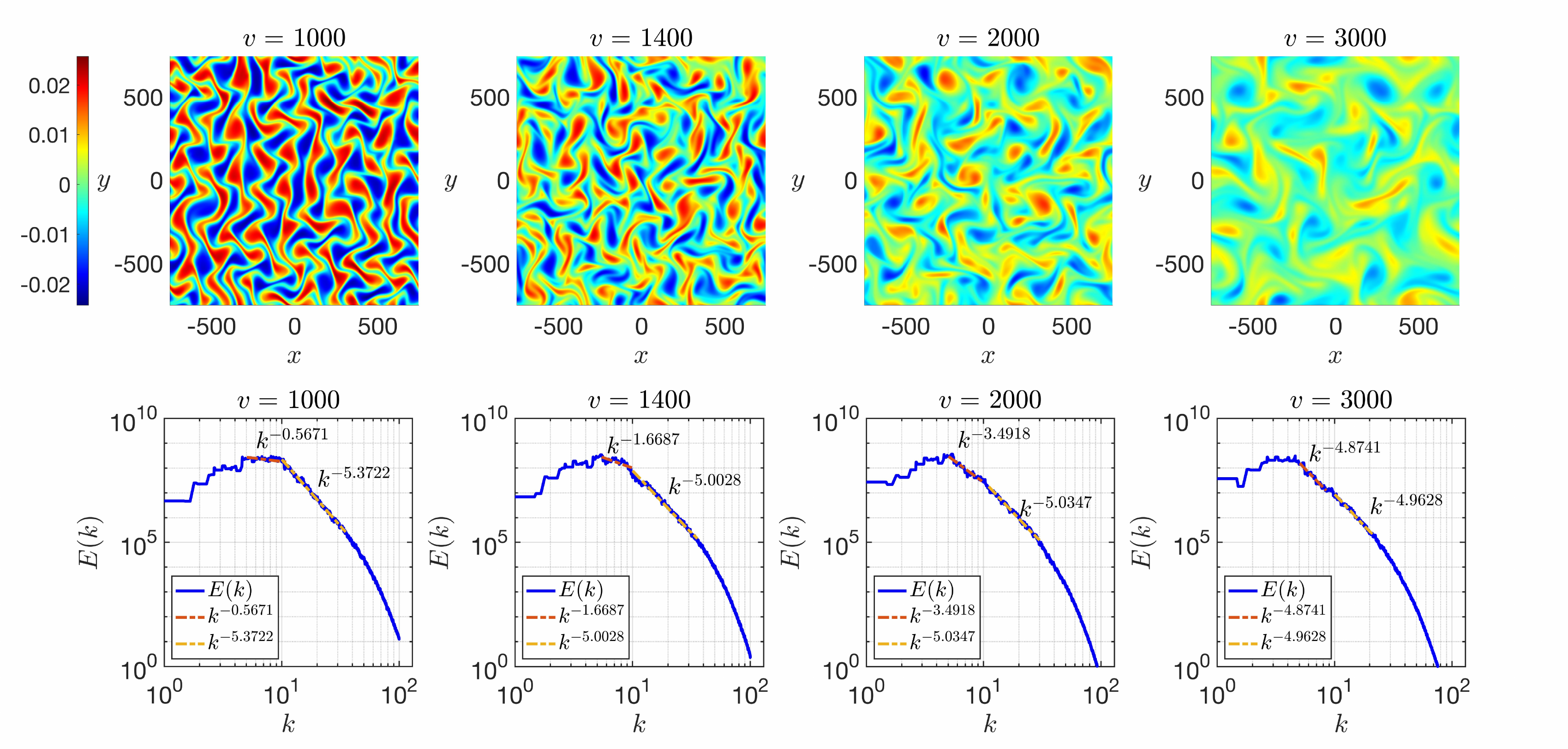}

\caption{The vorticity field $\omega=\partial_{x}u_{y}-\partial_{y}u_{x}$
of the boundary fluid at $v=1000,1400,2000,3000$.
The flow is transformed from an unstable shear flow to a homogeneous
and isotropic decaying turbulent flow where the inverse cascade is
manifestly shown from the first row profiles. The stage at $v=1400$
corresponds to the point where the velocity components $u_{x}$ and $u_{y}$ reach approximately
the same order of magnitude. The plots in the second row show the
corresponding energy spectrum with fitted scaling power in two
subranges $k\in\left(5,10\right)$ and $k\in\left(10,35\right)$.
\label{fig:decaying_turbulence_vorticity_and_total_Ek_clim_fixed}}
\end{figure*}

\begin{figure}
\centering
\includegraphics[scale=0.45]{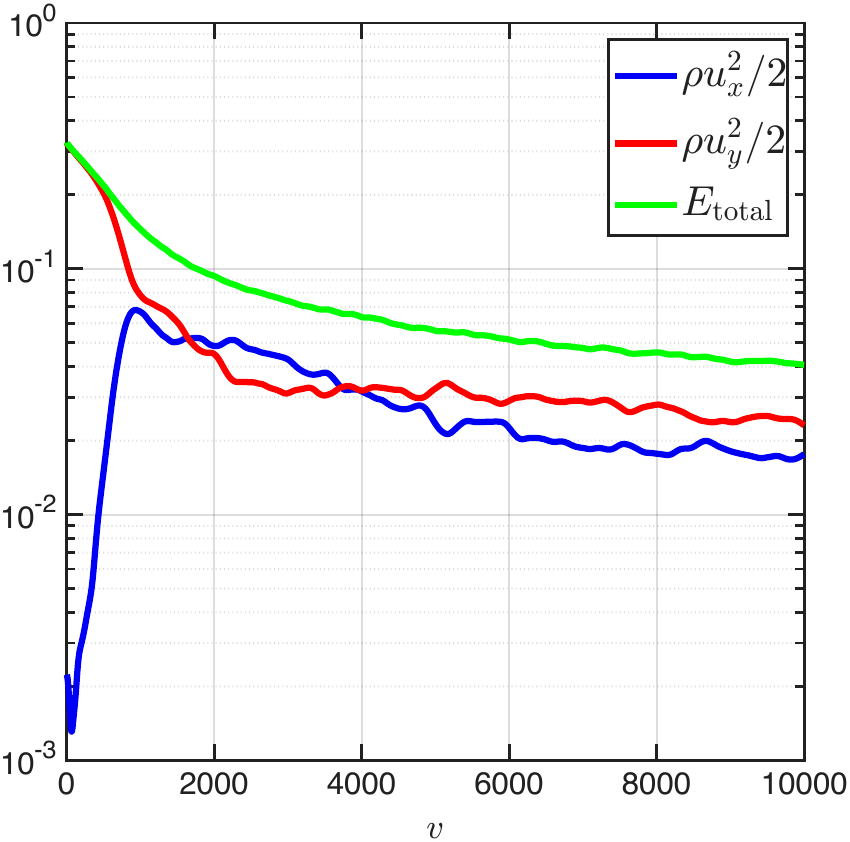}

\caption{Mean kinetic energy $E_{\mathrm{total}}=\frac{1}{2L^{2}}\protect\int d^{2}x\rho\bm{u}^{2}$
of the fluid and its $u_{x}$ and $u_{y}$ contributions from $v=0$
to $v=10000$. It shows that the shear flow transforms into turbulence
by two stages: while the total kinetic energy $E$ decreases, nonlinear
instability triggers $u_{x}$ to grow exponentially until it reaches
the same level as $u_{y}$ (at $v\approx 1400$) and then both $u_{x}$ and $u_{y}$ decay
at a similar rate. \label{fig:mean_kinetic_total_energy_of_decaying_turbulence}}
\end{figure}

In this section, we shall present our numerical results of the freely
decaying turbulence where no driving force is imposed. The bulk geometry
is then described by the pure Einstein gravity, within which the corresponding
dual fluid undergoes a transition from a shear flow to a turbulent
flow.

\subsection{Initial Fluid Configuration}

We consider an unstable shear flow in a periodic box with equal size
$L_{x}=L_{y}=L$ as our initial configuration, which is also considered
in \citep{HolographicTurbulence2014,NumericalSolutionGravitational2014,IncompressibleNonRelativisticNavierStokes2009,TurbulentFlowsRelativistic2012,HolographicPathTurbulent2014,HolographicTurbulenceLarge2018,HolographicTurbulenceEinsteinGaussBonnet2019}.
The velocity field is given by
\begin{equation}
u_{x}\left(x,y\right)=\delta u_{x}\left(x,y\right),\quad u_{y}\left(x,y\right)=A_{y}\cos\left(Qx\right),\label{eq:decaying_init_fluid_config}
\end{equation}
where small perturbations 
\begin{equation}
\delta u_{x}=\delta A_{x}\sum_{k}c_{k}\cos\left(\bm{k}\cdot\bm{x}+\theta\right)\label{eq:shear_flow_perturbations}
\end{equation}
with random phases $\theta$ and amplitudes $c_{k}$ are added to
trigger the fluid's instability at high enough Reynolds number. This
initial fluid configuration is dual to a locally boosted black brane
with the same boost velocity as in \prettyref{eq:decaying_init_fluid_config},
from which the initial value for $h_{ij}$ can be readily obtained \citep{LosingForwardMomentum2014,HolographicTurbulence2014,NumericalSolutionGravitational2014}.
Meanwhile, we impose the initial boundary conditions for $f,\xi,\eta$
from matching the boundary energy-momentum tensor \prettyref{eq:boundary_Tab}
with the zeroth-order derivative expansion of the stress tensor for a conformal
fluid \citep{TurbulentFlowsRelativistic2012,GravityHydrodynamicsLectures2009}:
\begin{equation}
T^{\mu\nu}=\alpha T^{3}\left(3u^{\mu}u^{\nu}+g^{\mu\nu}\right)+O\left(\nabla\right),
\end{equation}
where $\alpha$ is a dimensionless normalization constant. The temperature
$T$ is related to the black hole horizon by $T=4\pi/\left(3z_{H}\right)$.
For a fixed characteristic length $L$, the higher-order derivative
terms are suppressed when $LT\gg1$, thus nonlinear advection dominates
over viscous effects. In practice, we set the box size $L=1500$,
the wavenumber $Q=20\pi/L$, fluid energy density and pressure $\rho=2P=2$
(i.e. $\alpha=1$). The initial velocity is chosen in the relativistic regime where amplitude $A_{y}=0.8$ and $\delta A_{x}$ is adjusted such that $\vert\delta u_{x}|_{\mathrm{max}}=0.2$.

\subsection{Vorticity and the Scaling Law}

The velocity field of the fluid is defined in the Landau frame, as
ambiguities in the definition of velocity arise in the relativistic
regime \citep{GravityHydrodynamicsLectures2009,LecturesHydrodynamicFluctuations2012}. In this frame, the
velocity field $u_{L}^{\mu}$ is obtained from the time-like eigenvector
of the energy-momentum tensor,
\begin{equation}
T_{\,\nu}^{\mu}u_{L}^{\nu}=-\rho u_{L}^{\mu}
\end{equation}
where $\rho$ is the local energy density of the fluid. 

As the system
evolves, the nonlinear advection dominates over the viscous dissipation,
during which the perturbations are amplified by the instability, and the
resulting dynamics break the initial translational symmetry along
the $y$ direction and eventually produce a turbulent flow. To illustrate it, the boundary vorticity field 
\begin{equation}
\omega=\partial_{x}u_{y}-\partial_{y}u_{x},\label{eq:def_vorticity},
\end{equation}
which describes the rotational motion of the fluid, is plotted. Four
profiles at different times are shown in Figure \ref{fig:decaying_turbulence_vorticity_and_total_Ek_clim_fixed}.
It clearly demonstrates the fluid's motion consists of many small
clockwise (blue) and counterclockwise (red) vortices. Those vortices
with the same rotation split and merge and gradually grow into bigger
ones, consequently reducing the total number of vortices and indicating
an obvious behavior of two-dimensional turbulent inverse cascade.
At late times, the system evolves into a state characterized by a
few pairs of very slowly moving large-scale coherent vortices with
opposite direction of rotation, as also observed in the incompressible
Navier-Stokes turbulence \citep{EmergenceIsolatedCoherent1984,DynamicsFreelyDecaying1988}.

To quantify the turbulence obtained from the gravitational evolution,
the energy spectrum of the turbulence is transformed in terms of
wavenumber $k$,
\begin{equation}
E\left(k\right)=\frac{1}{2}\partial_{k}\int_{\left|\bm{k}\right|<k}d^{2}\bm{k}\left|\widetilde{\bm{w}}\right|^{2},\label{eq:def_Ek}
\end{equation}
where the tilde denotes the Fourier transformation and $\bm{w}=\sqrt{\rho}\bm{u}$.
Because the fluid energy density varies and the velocity field is
not divergence free, the system does not correspond to an incompressible
fluid. Following the analysis of the compressible fluid \citep{EnergySpectralDynamics1990,CascadeKineticEnergy2013,ScalingLawsCompressible2017,UniversalityScalingHomogeneous2020},
we therefore multiply the fluid velocity with a density factor $\sqrt{\rho}$.
The energy spectrum peaks at $k=10$ since initially
we choose $Q=20\pi/L$ in \prettyref{eq:decaying_init_fluid_config}.
It subsequently grows progressively to lower wavenumbers and eventually
dominates there as the flow evolves. Several profiles are listed in the second
row of Figure \ref{fig:decaying_turbulence_vorticity_and_total_Ek_clim_fixed}.
These energy spectra are fitted with two subranges of $k$ in units
of $2\pi/L$. It seems that an approximate Kolmogorov $k^{-5/3}$
scaling emerges at around $v=1400$ between $k\in\left(5,10\right)$.
In addition, a scaling of $k^{-5}$ appears around
$k\in\left(10,35\right)$. These observations are consistent with
those in \citep{HolographicTurbulence2014}.

However, this approximate $k^{-5/3}$ observed in the present setup
may not provide a reliable estimate of the fractal dimension, as the
scaling power is transient and sensitive to the fitting range. As
shown in Figure \ref{fig:decaying_turbulence_vorticity_and_total_Ek_clim_fixed},
this inertial range is narrow where few scales are covered, and later time evolution
shows this $k^{-5/3}$ scaling soon shifts to around $k^{-5}$
as a result of coherent vortex formation \citep{NonrobustnessTwoDimensionalTurbulent2007,EmergenceIsolatedCoherent1984}.
Also the range of $k^{-5}$ scaling becomes shorter and moves
slowly to the large scales because of the inverse energy cascade. To extend both the
lifetime of the energy spectrum and the width of its inertial range, two straightforward approaches may be employed.
On the one hand, energy can be externally and consistently injected
into the fluid, since the observed results indicate the turbulence
is not in a steady state and the $k^{-5/3}$ scaling appears only
at early times. This can also be explicitly seen in Figure \ref{fig:mean_kinetic_total_energy_of_decaying_turbulence}
where the mean kinetic energy of the turbulent flow,
\begin{equation}
E=\frac{1}{2L^{2}}\int d^{2}x\bm{w}^{2},\label{eq:def_mean_Ek}
\end{equation}
is decomposed into two components in order to characterize the respective
contributions from $u_{x}$ and $u_{y}$. Specifically, each component
is defined as $E_{i}=\frac{1}{2L^{2}}\int d^{2}x\left(\sqrt{\rho}u_{i}\right)^{2},i=x,y$.
In Figure \ref{fig:mean_kinetic_total_energy_of_decaying_turbulence},
the $u_{x}$ component of the kinetic energy grows exponentially until
it reaches a magnitude comparable to that of $u_{y}$ around $v=1400$.
Afterward, both $u_{x}$ and $u_{y}$ decay at rates similar to that
of the total kinetic energy, and a clear $k^{-5/3}$ scaling becomes
increasingly difficult to resolve. On the other hand, one may extensively enlarge
the box size $L$ and numerical resolution. This could allow
a much higher Reynolds number and cover more scales, but such direct
simulations typically demand greater computational cost than external energy driving does.

\section{Forced Turbulence\label{sec:ForcedTurb} }

\begin{figure*}
\centering
\includegraphics[scale=0.5]{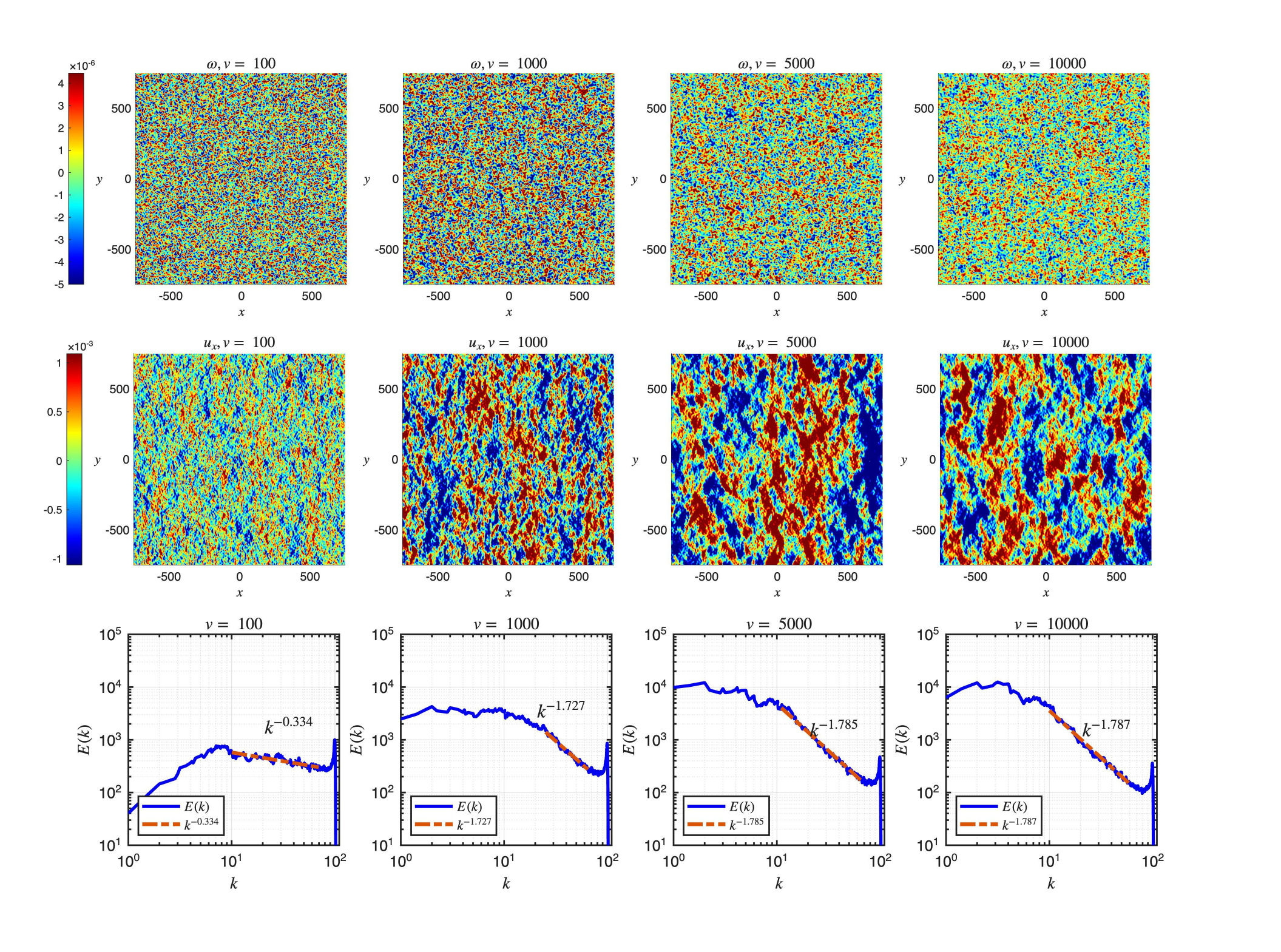}

\caption{Vorticity field $\omega$, velocity field component $u_{x}$ and the energy spectrum
$E\left(k\right)$ of the driven turbulence at $v=100, 1000, 5000, 10000$. The vorticity field (top
row) is manifestly homogeneous and isotropic. Vortices grow from the driving
scale $k_{f}$ to the largest scales around $k=10$ which agree
with the energy spectrum. The large-scale structures observed in the
velocity component $u_{x}$ (middle row) are similar to those in $u_{y}$.
A narrow peak in the energy spectrum $E\left(k\right)$ (bottom row)
appears around the driving scale $k_{f}=100$ where energy is injected
and transferred into large scales. \label{fig:driven_turb_vort_ux_Ek}}
\end{figure*}

In this section, we introduce a random and periodic driving force to evolve a driven turbulent flow and analyze its spectrum.

\subsection{An External Driving Force}

As discussed previously, to realize a steady turbulent flow, energy
should be continuously injected into the system to balance the dissipation
produced by the viscosity, which necessitates an external driving force.
We then introduce a massive scalar field $\phi$ in the bulk spacetime
as described in Section \ref{sec:TheDualGravModel}. Similar considerations
can be found in \citep{ForcedFluidDynamics2009,PeriodicallyDrivenAdS2013,ForcedFluidDynamics2014,DrivenHolographicCFTs2015}.
The boundary fluid is subject to the equation of motion \prettyref{eq:bdry_fluid_EOM},
where the nontrivial boundary value of the scalar field together with
its response play the role of an external force. We source $\phi_{1}$ and interpolate its
value through the following equation
\begin{eqnarray}
d_{v}\phi_{1} & = & -\frac{\pi}{2\Delta v}\sin\left(\frac{\pi}{2}\frac{v-v_{1}}{\Delta v}\right)F\left(v_{1},\bm{x}\right)\nonumber \\
 &  & +\frac{\pi}{2\Delta v}\cos\left(\frac{\pi}{2}\frac{v-v_{2}}{\Delta v}\right)F\left(v_{2},\bm{x}\right),\label{eq:choice_random_driving}
\end{eqnarray}
where $F\left(v,\bm{x}\right)$ is given by
\begin{eqnarray}
F\left(v,\bm{x}\right) & = & \mathcal{A}\sum_{i=0}^{n}c_{i}\left(v\right)\cos\left(\frac{2\pi}{L}\bm{k}\cdot\bm{x}+\theta_{i}\left(v\right)\right).
\end{eqnarray}
Here, $\mathcal{A}$ is the amplitude and $n$ is the number of random
modes. The random amplitudes $c_{i}\left(v\right)$ are drawn from
a normal distribution with zero mean and variance $\Delta v$ while
random phases $\theta_{i}\left(v\right)$ are drawn from a uniform distribution.
Both of these random variables update every $\Delta v$.  The source is then constructed by periodic interpolation between randomly drawn values, yielding a deterministic driver within each interval, similar in spirit to \citep{DrivenBlackHoles2021}. 
We set
$n=200$, $\mathcal{A}=0.02$, $L=1500$ and $\Delta v=20\delta v$,
where $\delta v$ is the numerical time step. The driving force is
imposed around a band-limited ring in momentum space at $k_{f}\pm\delta k$
with $k_{f}\simeq 100$ and $\delta k=1$. This driving scale is
selected near the minimal numerical resolution, which enables us
to maximize the exploration of the inverse cascade and widen the
inertial range as much as possible. The initial gravity configuration
is chosen as a Schwarzschild-$\mathrm{AdS}_{4}$ black hole,

\begin{equation}
ds^{2}=\frac{\ell^{2}}{z^{2}}\big[-\big(1-\frac{z^{3}}{z_{h}^{3}}\big)dv^{2}-2dvdz+dx^{2}+dy^{2}\big],
\end{equation}
which corresponds to a conformal fluid in a thermal state with universal
shear viscosity over entropy ratio \citep{ViscosityStronglyInteracting2005}.

\subsection{The Energy Spectrum of Forced Turbulence}

\begin{figure}
\centering
\includegraphics[scale=0.45]{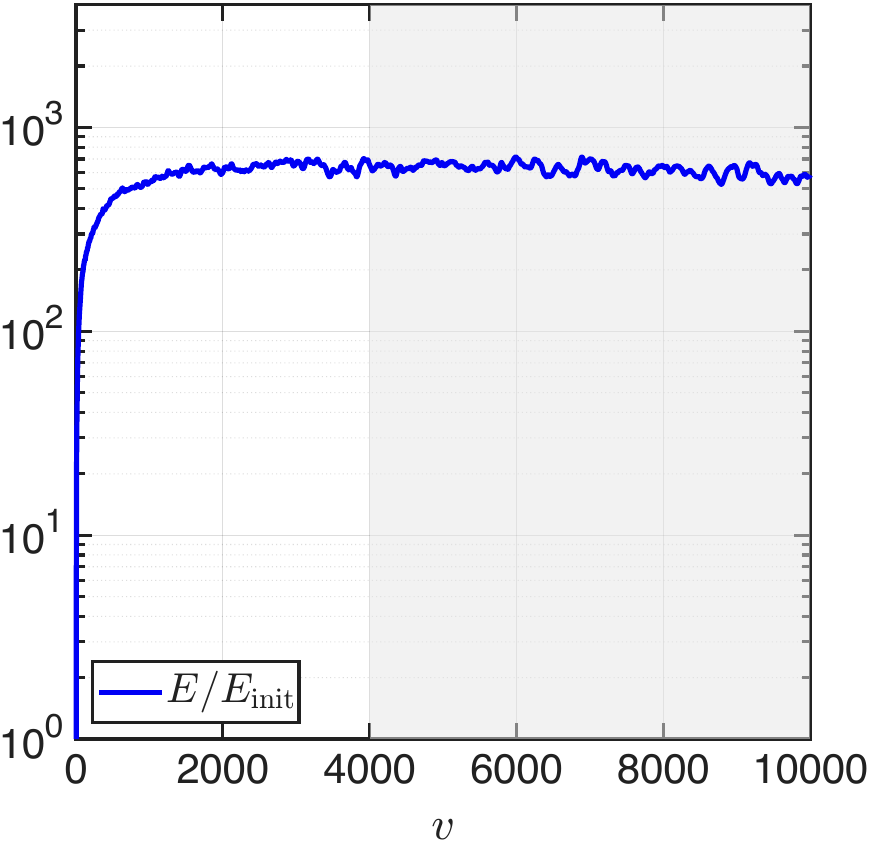}

\caption{Mean kinetic energy \eqref{eq:def_mean_Ek} of the driven turbulence.
It is normalized by the value at $v=1$ since it vanishes at the initial time.
It rapidly rises from a small value and then fluctuates around a
nearly constant value. Shaded area shows the range used in the time average of the scaling powers.\label{fig:driven_turb_mean_total_Ek}}
\end{figure}

\begin{figure}
\centering
\includegraphics[scale=0.40]{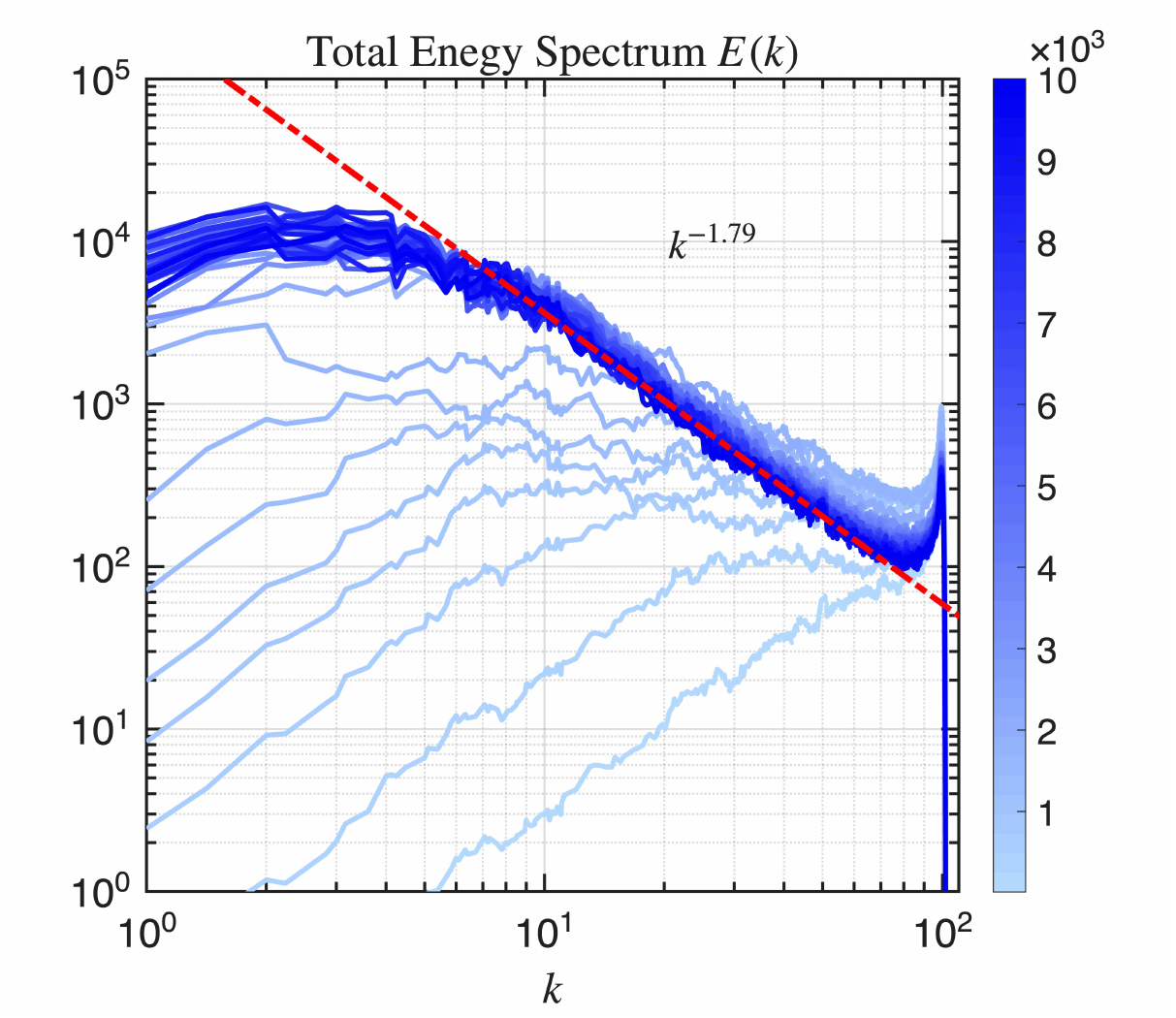}

\caption{Total energy spectrum of the driven turbulence. Lines with deeper colors
correspond to later times. The figure clearly demonstrates an inverse
energy cascade in which energy is transferred from large to small
scales. At late times, a power law scaling emerges within the inertial
range. A time-averaged fit over the range of each scaling around $k\in\left(10,65\right)$
from $v=4000$ to $v=10000$ yields a scaling exponent of $-1.79\pm0.03$,
indicated by the red dash-dotted line.\label{fig:driven_turb_Ek_v}}
\end{figure}

\begin{figure}
\centering
\includegraphics[scale=0.45]{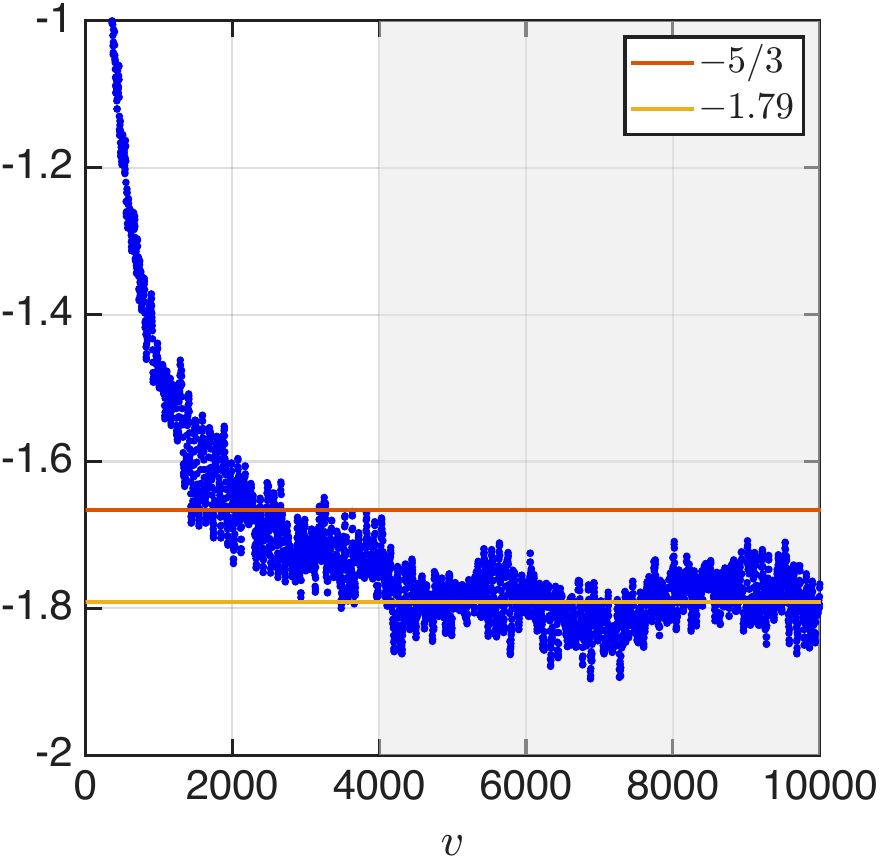}

\caption{Fitted scaling power law exponents in the inertial range $k\in\left(10,65\right)$. Because the energy spectrum is still developing at early times, only data from $v=4000$ to $v=10000$ is used to take a time average, yielding a power-law exponent of $-1.79\pm0.03$ (green line). A reference line with a value $-5/3$ (red line) indicates Kolmogorov's scaling.\label{fig:driven_scaling_powers_v=00003D4e3_1e4}}
\end{figure}

Once the driving force is turned on, the fluid is rapidly excited
into a turbulent state. Figure \ref{fig:driven_turb_vort_ux_Ek} shows
the numerical evolution of the vorticity \prettyref{eq:def_vorticity},
velocity component $u_{x}$ and total kinetic energy spectrum \prettyref{eq:def_Ek}
at four different representative times. It clearly reveals
the homogeneous and isotropic character of the turbulent flow, the development of small-scale velocity fluctuations into large-scale structures and the growth of the energy spectrum into small wavenumbers. Also, the maximum velocity of the fluid is $\bm{u}\sim10^{-3}\ll1$, which lies in the non-relativistic regime.

Before discussing the energy spectrum, we first see how the kinetic
energy of the fluid changes. Figure \ref{fig:driven_turb_mean_total_Ek} shows the evolution of the mean kinetic energy, while Figure \ref{fig:driven_turb_Ek_v} presents the corresponding energy spectrum of the driven turbulence. In contrast to that in the decaying case
(see Figure \ref{fig:mean_kinetic_total_energy_of_decaying_turbulence}),
it exhibits a rapid initial increase and then fluctuates about an
approximately constant value. This indicates that the turbulent flow
reaches a quasi-steady state. We term it quasi-steady because the
fluid is confined in a finite domain and no large-scale friction is
provided, which allows energy to accumulate at the system scale $L$.
Over sufficiently long time evolution, this accumulation may lead
to energy condensation and the emergence of coherent vortices \citep{DynamicsEnergyCondensation2007,DrivenBlackHoles2021}.
In our simulations, however, the system was not evolved for such a
prolonged period, so no energy condensation is observed. The largest
scale of the vortices is about $k=10$, which gives a wavelength $\lambda\sim2\pi/k\sim150$
as can be seen in Figure \ref{fig:driven_turb_vort_ux_Ek}.

The energy spectrum exhibits a distinct peak at the forcing
wavenumber $k_{f}$, which lies close to the dissipative scale. A
clear scaling behavior is observed in the intermediate range $k\in\left(10,65\right),$
between the driving scale $2\pi/k_{f}$ and the largest system scale $L$;
see Figure \ref{fig:driven_turb_Ek_v}. Based on the growth of the total kinetic energy and the energy
spectrum (see Figure \ref{fig:driven_turb_mean_total_Ek} and Figure \ref{fig:driven_turb_Ek_v}), a time average of the
fitted scaling exponents is taken over $v=4000$ to $v=10000$, yielding an energy spectrum scaling as
\begin{equation}
E\left(k\right)\sim k^{-1.79\pm0.03}.\label{eq:driven_total_Eks_averaged}
\end{equation}
As shown in Figure \ref{fig:driven_scaling_powers_v=00003D4e3_1e4}, the fitted scaling power fluctuates around a constant level in the range $k\in(10, 65)$ at late times and those exponents are smaller than $-5/3$. 

\begin{figure}
\centering
\includegraphics[scale=0.4]{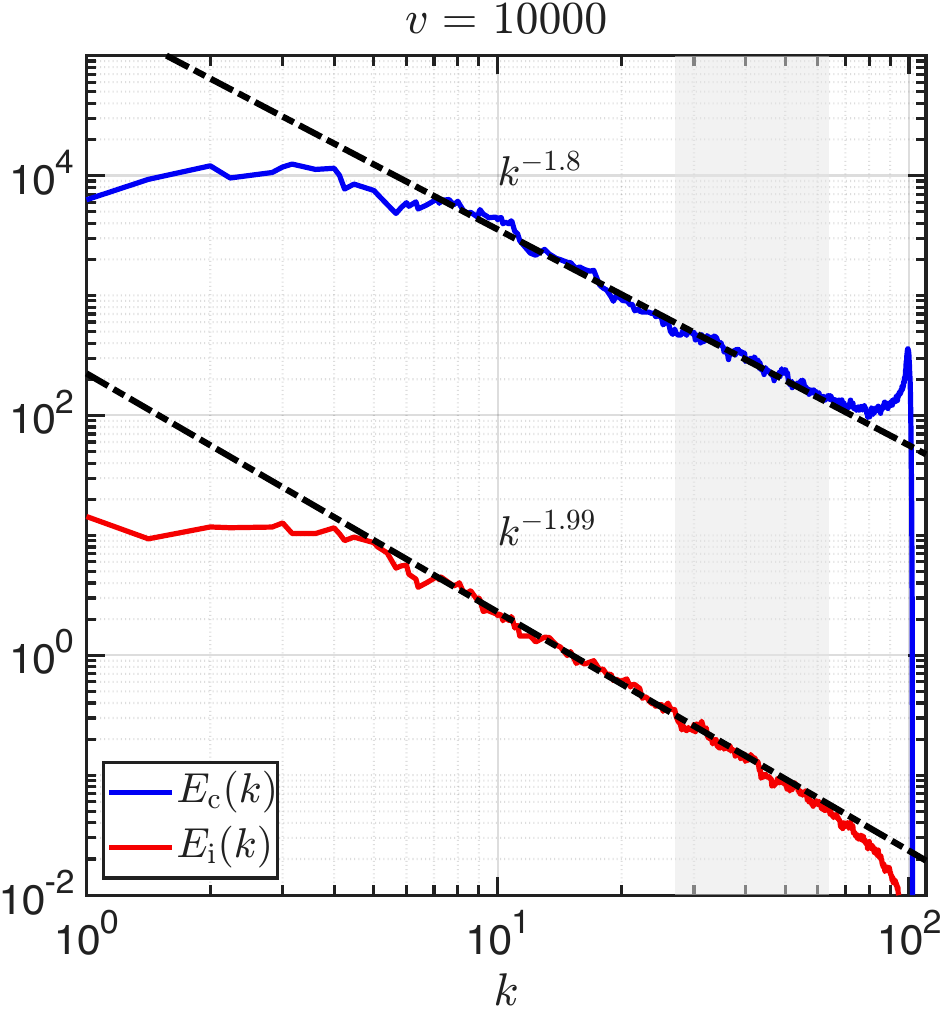}

\caption{A representative energy spectrum decomposed into compressible and
incompressible components. Two power-laws \eqref{eq:driven_incom_comp_Eks_time_averaged} are obtained from the time average. The incompressible component $E_{\mathrm{i}}\left(k\right)$ does not
show any feature at $k_{f}$. To avoid clutter, only the profile at
$v=10000$ is shown as the late time spectrum exhibits similar behaviors.
\label{fig:driven_profiles_of_com_incomp_EKs}}
\end{figure}

Such a deviation from the well-known Kolmogorov $k^{-5/3}$ scaling is to be expected.
The form of the external force \eqref{eq:bdry_fluid_EOM} indicates
that the fluid is not driven in an incompressible way. Therefore, although the
fluid is non-relativistic, it is not incompressible. By decomposing
the fluid velocity into the solenoidal (incompressible) and irrotational
(compressible) components 
\begin{equation}
\bm{u}=\bm{u}_{i}+\bm{u}_{c},\quad\nabla\bm{\cdot}\bm{u}_{i}=0,\quad\nabla\times\bm{u}_{c}=0,
\end{equation}
we find that the compressible component of the kinetic energy dominates;
see a representative profile in Figure \ref{fig:driven_profiles_of_com_incomp_EKs}.
A time-averaged power-law fit from $v=4000$ to $v=10000$ yields
\begin{equation}
E_{\mathrm{i}}\sim k^{-1.99\pm0.03},\quad E_{\mathrm{c}}\sim k^{-1.80\pm0.03}.\label{eq:driven_incom_comp_Eks_time_averaged}
\end{equation}
where $E_{\mathrm{i}}$ and $E_{\mathrm{c}}$ are fitted in the range $k\in\left(10,50\right)$
and $k\in\left(10,65\right)$ respectively. The scaling law \eqref{eq:driven_incom_comp_Eks_time_averaged} for
each component agrees well with the numerical experiments of the two-dimensional Navier-Stokes
compressible turbulence \citep{EnergyTransferSpectra2019} within
the range $k<k_{f}$. In \citep{EnergyTransferSpectra2019}, the compressible
turbulence is driven in a divergence-free way, and the power law of the total energy spectrum is close to that of the incompressible part. In our case, the compressible component dominates, thus the power law of the total energy spectrum scaling $k^{-1.79}$ is close to the compressible one $k^{-1.80}$. Despite the different ways of driving, the scaling law exponents of the incompressible and the compressible components are consistent and both are smaller than Kolmogorov's $k^{-5/3}$ scaling. Similar fitting exponents are also observed in \citep{HolographicTurbulenceRandom2024}. However, their inertial range is limited, which results in poorly constrained scaling exponents with large uncertainties. 

Even if the fluid is incompressible, the scaling may also deviate from Kolmogorov's $k^{-5/3}$ scaling. Previous two-dimensional incompressible Navier-Stokes numerical experiments \citep{NonrobustnessTwoDimensionalTurbulent2007} imply that the universality of Kolmogorov's $k^{-5/3}$ is not
robust and depends on the resolution of scales below the driving scale.
When these scales are well resolved, the energy spectrum exhibits a $k^{-2}$ scaling; otherwise, it exhibits a $k^{-5/3}$ scaling. Simulations of ideal conformal fluid with numerical dissipation \citep{FractalDimensionTurbulent2017} seem to find such an additional scaling. Also, the scaling exponents may depend on the spatial dimensions and the interactions of the incompressible and the compressible components. Three-dimensional simulations and analyses of compressible Navier-Stokes turbulence \citep{ConservativeCascadeKinetic2012,CascadeKineticEnergy2013,ScalingLawsCompressible2017} demonstrate that the total energy spectrum follows a $k^{-5/3}$ scaling, while the compressible component exhibits a $k^{-2}$ scaling at moderate Mach number which is defined as the ratio of fluid velocity over the speed of sound.  As Mach number increases, the interactions between the compressible and incompressible components will become stronger and hence influence the scaling behaviors. In our case, the velocity of the flow is much less than the speed of sound $c_{s} = 1/\sqrt{2}$ in the two-dimensional conformal boundary, which results in a small Mach number. 

\section{Fractal Dimension of Turbulent Black Holes\label{sec:fractal_dim_turbulent_BH}}

\begin{figure}
\centering
\includegraphics[scale=0.43]{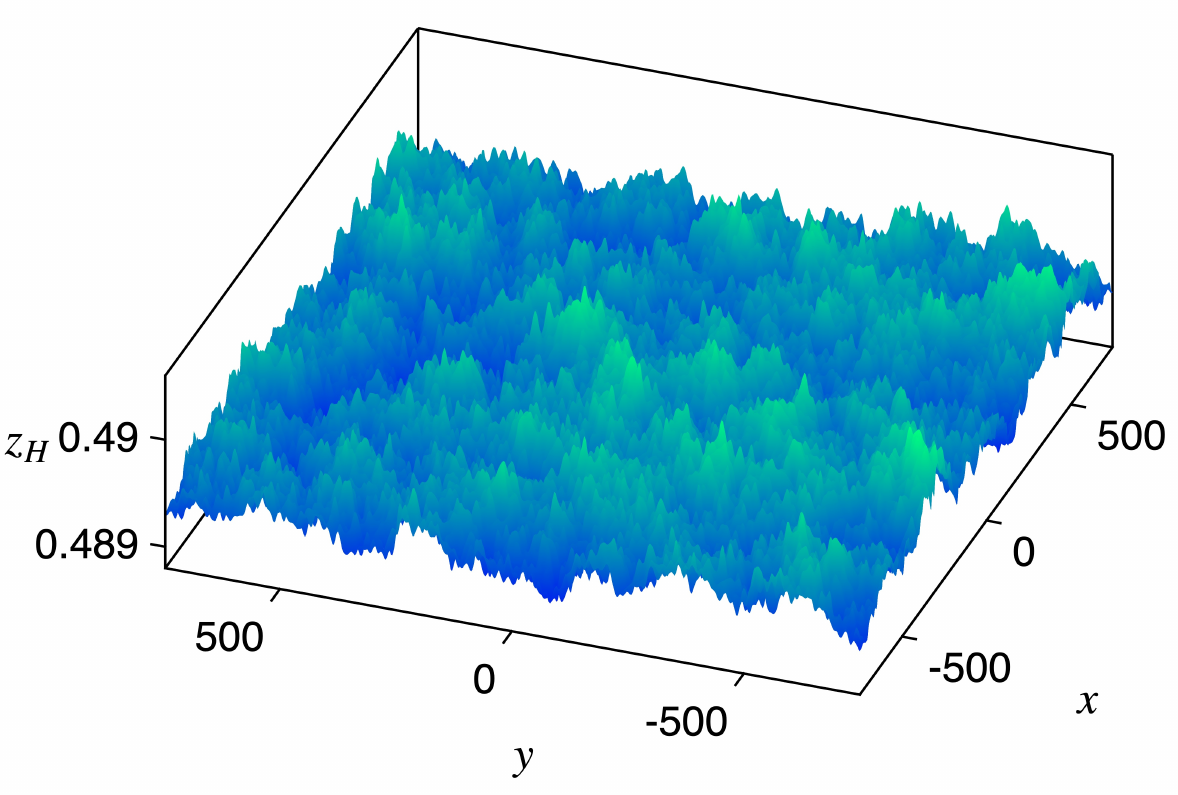}

\caption{Fractal structure of the apparent horizon $z_{H}$ at $v=10000$. \label{fig:driven_zHs_v=10000}}
\end{figure}

\begin{figure}
\centering
\includegraphics[scale=0.4]{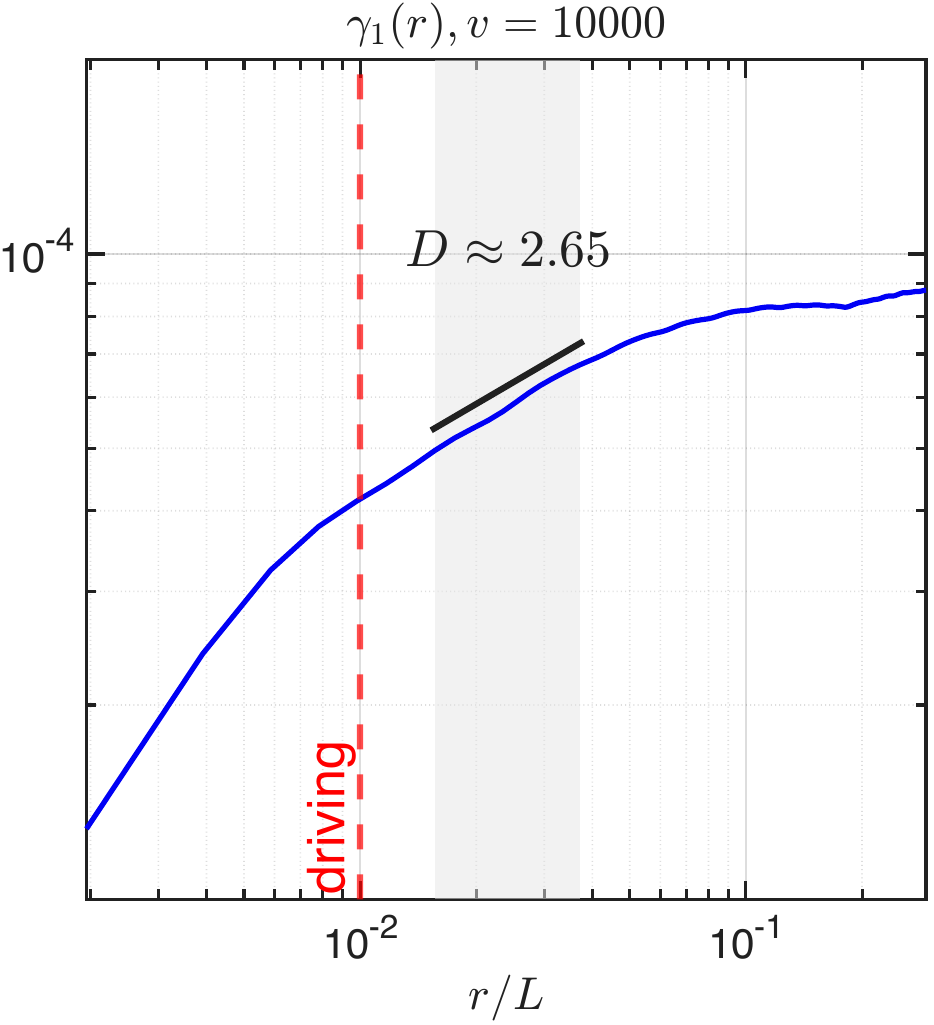}

\caption{An averaged madogram for transections of $z_{H}$ at $v=10000.$ The
range used for the linear fit is shaded and corresponds to the range
in Figure \ref{fig:driven_profiles_of_com_incomp_EKs}. The behaviors
of the madogram are similar to those at late time evolutions. \label{fig:driven_madogram_profile}}
\end{figure}

As already mentioned, one of our main motivations is to obtain a more reliable estimate of the
fractal dimension of the turbulent black holes. This geometric object
may behave as a characteristic property of black holes. Since the
proposed definition of fractal dimension through the Riemann sum in
\citep{HolographicTurbulence2014} does not derive a reasonable result
and is not even reliable in one-dimensional curves embedded in the Euclidean
plane as argued in \citep{FractalDimensionTurbulent2017}, further
investigations are needed. In \citep{FractalDimensionTurbulent2017},
simulations of weakly compressible ideal conformal fluid on the boundary were
taken \citep{NumericalMeasurementsScaling2017}, and the black hole
horizon was constructed from the fluid/gravity duality up to the ideal order derivative approximation. 
Then using a madogram estimator \citep{EstimatorsFractalDimension2012}, two fractal dimensions $D=2.645\left(5\right)$ and $D=2.584(1)$ were estimated, corresponding to $k^{-5/3}$ and $k^{-2}$ scaling law respectively. However, since dissipative effects are crucial 
in the theories of hydrodynamics while an ideal fluid does not dissipate, a numerical dissipation term \citep{NumericalMeasurementsScaling2017} was added \citep{FractalDimensionTurbulent2017}. 
To avoid this, we perform a nonlinear bulk evolution dual to the boundary fluid with the correct viscosity. Then we follow the definition of the fractal dimension suggested in \citep{FractalDimensionTurbulent2017}, and make an estimate of the fractal dimension for the driven turbulent black hole we obtained.

First, the madogram for a one-dimensional curve $f\left(x\right)$
is defined as 
\begin{equation}
\gamma_{1}\left(r\right)=\frac{1}{2}\left\langle \left|f\left(x+r\right)-f\left(x\right)\right|\right\rangle ,
\end{equation}
where the bracket denotes the spatial average over $x$ and $r$ is the separation along $x$ between any two points on the curve. Then such a fractal dimension $D$ is obtained
through the expected scaling $\gamma_{1}(r)\propto r^{2-D}$. Figure \ref{fig:driven_zHs_v=10000}
shows the fluctuations of the apparent horizon $z_{H}$ and its fractal
structure. So for this turbulent black hole, we implement the madogram
estimation to each one-dimensional transect along each spatial direction
of the apparent horizon's location $z_{H}\left(v,x,y\right)$ at each
time. After taking a spatial average of those madograms, a fractal
dimension $D_{\mathrm{transect}}$ for the transects can be found.
Then, the fractal dimension of the black hole horizon at each time
is estimated through $D=D_{\mathrm{transect}}+1$; see Figure \ref{fig:driven_madogram_profile}
for a profile at $v=10000$. Finally, a time average from the shaded
range in Figure \ref{fig:driven_turb_mean_total_Ek} yields 
\begin{equation}
D=2.65\pm0.02,\label{eq:turb_BH_fractal_D}
\end{equation}
which corresponds to the total energy spectrum scaling $k^{-1.79}$ of the boundary turbulence.

Our measurement of the turbulent black hole's fractal dimension \eqref{eq:turb_BH_fractal_D} shows a surprisingly good agreement with the one in \citep{FractalDimensionTurbulent2017}, while there are two important distinctions. First, our result comes with direct nonlinear evolution of black hole dynamics where the viscous boundary fluid is simultaneously and naturally evolved. The dissipative effects of the fluid dual to the black hole are correctly considered. Moreover, this nonlinear bulk evolution goes beyond a truncation of derivative expansion. But in \citep{FractalDimensionTurbulent2017}, 
a derivative expansion of the fluid was carried out to ideal order, with the dissipative effects introduced due to numerical considerations. Second, the corresponding kinetic energy spectrum is different. The scaling exponent of turbulent kinetic energy spectrum is smaller than Kolmogorov's $-5/3$. The driving force of \citep{FractalDimensionTurbulent2017} is divergence-free, which suppresses direct excitation of compressive modes and leads to flow behavior close to incompressible dynamics in the non-relativistic regime. Our driving force in \prettyref{eq:bdry_fluid_EOM} is not divergence-free and therefore generally contains both compressive and solenoidal components, resulting in a turbulent flow dominated by compressive energy. Nevertheless, the agreement of the results suggests an interesting universality of the fractal dimension of the turbulent black hole horizon. The origin of such a universality still deserves future investigations.

\section{Summary and Discussion\label{sec:SumDiscussion}}

The present study reveals that when the holographic turbulence is driven by a scalar source, the scaling power-law of the total energy
spectrum exists and behaves as $k^{-1.79\pm0.03}$, which is different from Kolmogorov's
$k^{-5/3}$ scaling for the incompressible and non-relativistic fluid. By decomposing
the spectrum into the compressible and the incompressible components, they
scale as $k^{-1.80\pm0.03}$ and $k^{-1.99\pm0.03}$ respectively.
Our system captures the non-relativistic limit, while the dual flow is not incompressible. Our results first clearly demonstrate that the kinetic energy spectrum scaling exponent of the quasi-steady holographic turbulence is not previously suggested Kolmogorov's $k^{-5/3}$. Furthermore, we correctly treat the dissipative effects in the boundary turbulent fluid through the fully nonlinear evolutions of driven black hole dynamics instead of the numerical dissipative terms added \citep{FractalDimensionTurbulent2017}. Based on these, from the scaling law of the boundary turbulence which indicates the existence of some scale-independent quantities, we estimate the fractal dimension of the turbulent black hole's horizon $D\approx 2.65\pm 0.02$. Our fractal dimension stays below the topological dimension of the horizon and shows a good agreement with the previous estimate through simulations of the boundary ideal conformal fluid. We provide the first estimation of fractal structure from the fully nonlinear evolution of driven black hole dynamics with the best resolution and widest inertial range so far compared to the related works in the literature. To do so, we have managed to develop an efficient evolution scheme in the Bondi-Sachs formalism for $\mathrm{AdS_{4}}$ spacetime, which is also possible for higher dimensions like $\mathrm{AdS_{5}}$.

It is also instructive to compare our results with those obtained in the limit of large spacetime dimension, i.e. large-$D$ limit \citep{LargeLimitGeneral2013,LargeLimitEinsteins2020}, where the Einstein equations are simplified significantly and the black branes can be shown to effectively evolve as viscous fluids \citep{EffectiveTheoryBlack2015,EvolutionEndPoint2015,HydroelasticComplementarityBlack2016,LargeHolographyMetric2018}. Previous works \citep{HolographicTurbulenceLarge2018,HolographicTurbulenceEinsteinGaussBonnet2019} simulated the decaying turbulence in the large-$D$ limit while no Kolmogorov's $-5/3$ is observed in two spatial dimensions. This is consistent with our observation that decaying turbulence in two dimensions does not sustain a clear $k^{-5/3}$ scaling. In \citep{DrivenBlackHoles2021}, by applying a divergence-free force, the obtained quasi-steady turbulence exhibits a clear Kolmogorov's $k^{-5/3}$ scaling law in the large-$D$ limit. This difference may stem from the fact that their driving directly sources the vorticity of the fluid and therefore the obtained turbulence is close to incompressible or weakly compressible, while our flow is dominated by the compressible component. It is interesting to see 
whether the large-$D$ framework with a non-divergence-free source would also yield a compressible-dominated spectrum as in our case. Furthermore, it is also intriguing to see the fractal dimension of the turbulent black brane in the large-$D$ limit.

There are many further directions to be investigated. First, it would be direct and interesting to increase the numerical resolution such that the driving scale is way larger than the smallest scale and to see whether the scaling law changes or not, since there are some indications possibly from the incompressible Navier-Stokes equations \citep{NonrobustnessTwoDimensionalTurbulent2007} and the ideal boundary conformal fluid \citep{FractalDimensionTurbulent2017}. One may also change the way of driving to achieve that the incompressible energy of the turbulence dominates, from which the incompressible condition is approximately satisfied. Second, as found in \citep{PeriodicallyDrivenAdS2013, DrivenHolographicCFTs2015},
there exist different phase structures of the CFTs dual to the Einstein-scalar system. One may extensively change the driving periodicity
and its strength to see how the scaling depends on the phase structures.
One can also add the interactions for the scalar field. This may lead
to a phase transition of the turbulent black hole. Finally, as discussed in \citep{FractalDimensionTurbulent2017}, since the fractal dimension of the turbulent black hole is a geometric quantity, its definition should be covariantly defined and be further investigated.

\begin{acknowledgments}
YT is grateful to Hong Liu for insightful discussions on the fractal dimension of the turbulent horizon during his visit to MIT. JD would also like to thank Amos Yarom, Sebastian Waeber, René Meyer, Bin Sun and Jiuling Wang for their helpful discussions. This work is partly supported by the National Natural Science Foundation of China with Grants No. 12075026, No. 12035016, No. 12361141825, and No. 12375058.
\end{acknowledgments}

\appendix
\section{Holographic Renormalization\label{sec:appendix_Holographic-Renormalization}}

Following the holographic renormalization processes \citep{StressTensorAntide1999,HolographicReconstructionSpacetime2001,HolographicRenormalization2002,LectureNotesHolographic2002,ThermodynamicsAsymptoticallyLocally2005},
the boundary energy-momentum tensor is given by 
\begin{equation}
\left\langle T^{\mu\nu}\right\rangle =\frac{2}{\sqrt{-\gamma}}\frac{\delta S}{\delta\gamma_{\mu\nu}}\label{eq:bdry_Tdd_variation_of_S}
\end{equation}
where the bulk action \prettyref{eq:einstein_scalar_action} is supplemented
with the Gibbons-Hawking-York term to make it a well-posed variational
problem and $S_{ct}$ to cancel out the divergence of the near-boundary
behavior

\begin{eqnarray}
S & = & S_{\mathrm{EH}}+S_{\mathrm{bdry}}+S_{\mathrm{ct}}\nonumber \\
 & = & \frac{1}{2\kappa^{2}}\int_{M}d^{4}x\sqrt{-g}\left(R-2\Lambda\right)\nonumber \\
 &  & -\frac{1}{\kappa^{2}}\int_{\partial M}d^{3}x\sqrt{-\gamma}K+\frac{1}{\kappa^{2}}S_{ct}\left(\gamma_{ab}\right).
\end{eqnarray}
Here, the action for the counter terms is chosen as
\begin{eqnarray}
S_{ct} & = & \int dx^{3}\sqrt{-\gamma}\left(-\frac{2}{\ell}\left(1+\frac{\ell^{2}}{4}\mathcal{R}\right)-\frac{1}{4\ell}\phi^{2}\right)
\end{eqnarray}
where $\mathcal{R}$ is the Ricci tensor corresponding to $\gamma_{\mu\nu}$
and the second term aims to cancel the divergence in the presence
of the scalar field. Then from \prettyref{eq:bdry_Tdd_variation_of_S},
we have
\begin{multline}
\left\langle T_{ab}\right\rangle =\frac{1}{\kappa^{2}}\lim_{\epsilon\rightarrow0}\bigg[\frac{\ell}{\epsilon}\bigg(K_{ab}-K\gamma_{ab}\\
-\frac{2}{\ell}\gamma_{ab}+\ell G_{ab}-\frac{1}{4\ell}\phi^{2}\gamma_{ab}\bigg)\bigg],\label{eq:boundary_Tab}
\end{multline}
where $G_{ab}$ is the Einstein equation on the boundary and vanishes automatically since the boundary geometry \prettyref{eq:spatial_metric}
we consider is conformally flat. The explicit form of the energy-momentum
tensor reads 
\begin{eqnarray}
\left\langle T_{vv}\right\rangle  & = & \ell^{2}\left(-f_{3}+\dfrac{1}{6}\phi_{1}\phi_{2}\right)\\
\left\langle T_{vx}\right\rangle  & = & \ell^{2}\left(-\dfrac{3}{2}\xi_{3}-\dfrac{1}{8}\phi_{1}\partial_{x}\phi_{1}\right)\\
\left\langle T_{vy}\right\rangle  & = & \ell^{2}\left(-\dfrac{3}{2}\eta_{3}-\dfrac{1}{8}\phi_{1}\partial_{y}\phi_{1}\right)\\
\left\langle T_{xx}\right\rangle  & = & \ell^{2}\left(-\dfrac{1}{2}f_{3}+\dfrac{3}{2}B_{3}+\dfrac{1}{3}\phi_{1}\phi_{2}-\dfrac{1}{4}\phi_{1}\partial_{v}\phi_{1}\right)\\
\left\langle T_{xy}\right\rangle  & = & \ell^{2}\left(\dfrac{3}{2}C_{3}\right)\\
\left\langle T_{yy}\right\rangle  & = & \ell^{2}\left(-\dfrac{1}{2}f_{3}-\dfrac{3}{2}B_{3}+\dfrac{1}{3}\phi_{1}\phi_{2}-\dfrac{1}{4}\phi_{1}\partial_{v}\phi_{1}\right)
\end{eqnarray}
The conservation of the boundary fluid is then 
\begin{eqnarray}
\partial_{v}f_{3} & = & \dfrac{1}{6}\partial_{v}\left(\phi_{1}\phi_{2}\right)-\dfrac{3}{2}\left(\partial_{x}\xi_{3}+\partial_{y}\eta_{3}\right)\nonumber \\
 &  & -\dfrac{1}{8}\left[\partial_{x}\left(\phi_{1}\partial_{x}\phi_{1}\right)+\partial_{y}\left(\phi_{1}\partial_{y}\phi_{1}\right)\right]\nonumber \\
 &  & +\left\langle O_{\phi}\right\rangle \partial_{v}\phi_{1},\label{eq:d1v_f3}\\
\partial_{v}\xi_{3} & = & -\dfrac{1}{12}\partial_{v}\left(\phi_{1}\partial_{x}\phi_{1}\right)-\partial_{y}C_{3}\nonumber \\
 &  & -\partial_{x}\left(-\frac{1}{3}f_{3}+B_{3}+\frac{2}{9}\phi_{1}\phi_{2}-\dfrac{1}{6}\phi_{1}\partial_{v}\phi_{1}\right)\nonumber \\
 &  & +\frac{2}{3}\left\langle O_{\phi}\right\rangle \partial_{x}\phi_{1},\label{eq:d1v_xi3}\\
\partial_{v}\eta_{3} & = & -\dfrac{1}{12}\partial_{v}\left(\phi_{1}\partial_{y}\phi_{1}\right)-\partial_{x}C_{3}\nonumber \\
 &  & -\partial_{y}\left(-\dfrac{1}{3}f_{3}-B_{3}+\dfrac{2}{9}\phi_{1}\phi_{2}-\dfrac{1}{6}\phi_{1}\partial_{v}\phi_{1}\right)\nonumber \\
 &  & +\frac{2}{3}\left\langle O_{\phi}\right\rangle \partial_{y}\phi_{1},\label{eq:d1v_eta3}
\end{eqnarray}
which are exactly components of the equation \prettyref{eq:bdry_fluid_EOM}.

\section{Details on Numerical Scheme\label{sec:appendix_numerical_details}}

\subsection{Bondi-Sachs Formalism}
In the seminal works of Bondi, Sachs and their collaborators \citep{GravitationalWavesGeneral1960,GravitationalWavesGeneral1962,GravitationalWavesGeneral1962a,GravitationalWavesGeneral1966},
the metric in the Bondi-Sachs gauge was first proposed to analyze the
gravitational radiation near null infinity in asymptotically flat
spacetime. This gauge is based on the outgoing null rays and demonstrates
the simple nested structure of the Einstein equation. For comprehensive
reviews on the Bondi-Sachs gauge and other possible gauge choices, see
\citep{CharacteristicEvolutionMatching2012,GeneralizedBondiSachsEquations2013,BondiSachsFormalism2016,AdvancedLecturesGeneral2019,SimulationsGravitationalCollapse2024,SimulationsGravitationalCollapse2024a}.
In the presence of non-zero cosmological constant, the Bondi-Sachs gauge
in AdS and dS spacetimes and integration schemes based on outgoing
null rays are discussed in \citep{AdS_4BondiGauge2019}. However,
as for black hole dynamics, a horizon penetrating scheme is needed,
thus null foliation based on ingoing null rays is more suitable; see
\citep{LosingForwardMomentum2014,NumericalSolutionGravitational2014,SpontaneousDeformationAdS2024}.
In \citep{LosingForwardMomentum2014}, a slightly modified Bondi-Sachs
gauge is chosen, while in \citep{NumericalSolutionGravitational2014}
and a series of subsequent works (see, e.g., \citep{HolographyCollidingGravitational2011,AffinenullMetricFormulation2013,HolographicTurbulence2014,CollidingShockWaves2015,NonlinearEvolutionAdS42019,HairyBlackResonators2022}),
the affine gauge is chosen. In this work, we adopt the original Bondi-Sachs
gauge as in \citep{SpontaneousDeformationAdS2024} which possesses
several appealing features that we discuss below.

The Bondi-Sachs metric \prettyref{eq:Bondi-Sachs-General-Metric}
is constructed by fixing the four gauge conditions
\begin{equation}
g^{vv}=0,\quad g^{vi}=0,\quad\partial_{r}\left(r^{-4}\det\gamma_{ij}\right)=0.\label{eq:BS_Gauge}
\end{equation} 
It provides the following appealing features for dynamical evolutions: 
\begin{itemize}
\item Under suitable initial and boundary conditions, the Einstein equation
\prettyref{eq:EFEs} reduces to a set of ordinary differential equations
(ODEs) on each null hypersurface which can be integrated explicitly.
Evolution between successive null hypersurfaces is governed by two evolution equations;
\item The choice of areal radius in the Bondi-Sachs gauge simplifies the Einstein equation. Notably, we find the evolution equations for $h_{ij}$
can be decoupled through a simple rotation \eqref{eq:SO_2_rotation}
in the absence of any spacetime symmetries. Furthermore, the radial
differential operators appearing in all equations involved in
the evolution scheme (see \eqref{eq:eom_chi}, \eqref{eq:eom_Pi},
\prettyref{eq:eom_f}, \prettyref{eq:eom_for_xi} and \eqref{eq:eom_PI_B_C_tilde}) are independent
of time $v$ and spatial coordinates $x^{i}$, which significantly
enhances numerical efficiency and especially enables us to accelerate
computations with GPUs;
\item A $z=\mathrm{const}$ hypersurface can be chosen as the inner boundary
for numerical computations and the spacetime inside this boundary
is excised; see Figure \ref{fig:schematic_evolution_fig}. On the
one hand, this is workable since the Bondi-Sachs gauge automatically
ensures that both the inner boundary and its tangent vector $\partial_{v}$
inside the black hole are space-like provided $f<0$.
Then all the boundary conditions can be imposed at the conformal boundary
$z=0$ thus they can be directly obtained from the boundary stress
tensor.
\item On the other hand, one can also fix the radial location of the apparent
horizon at a $z=\mathrm{const}$ hypersurface as in \citep{LosingForwardMomentum2014}.
This prevents the numerical domain from stepping too deeply into the
interior of black holes. However, such an implementation requires additional
auxiliary fields and one boundary condition should be placed at the
inner boundary. The value of this boundary condition is obtained from
the solution of a linear elliptic PDE at runtime which is typically
numerically challenging and resource intensive in two or higher spatial
dimensions. A similar situation also arises in the affine gauge \citep{NumericalSolutionGravitational2014}.
Suitable domain decomposition techniques \citep{ChebyshevFourierSpectral2001,NumericalSolutionGravitational2014,MultidomainSpectralMethod2003,NewPseudospectralCode2022}
may be viable and relatively enhance the numerical speed while it
increases code complexity in return. We do not fix the location
of the apparent horizon. Its location is determined independently
of the evolution and efficiently solved as described in Appendix
\ref{subsec:Locate-the-Apparent}. This avoids significant computational
cost in $\left(3+1\right)$-dimensional evolutions.
\end{itemize}
Although the black hole apparent horizon remains unfixed in our case
and may approach the interior singularity at very late
times, it can be mitigated by resetting the computational domain based
on the apparent horizon's location.

\subsection{Numerical Scheme \label{subsec:Numerical-Scheme}}
As mentioned above, the hierarchical structure in the Bondi-Sachs formalism allows us to solve the nonlinear Einstein equation through a sequence of ordinary differential equations if suitable initial conditions are
specified on an ingoing $v=\mathrm{const}$ null surface and boundary
conditions are imposed on a $z=\mathrm{const}$ hypersurface. Writing the
Einstein equation in trace-reversed form, the components $E_{zz}$,
$E_{zi}$, $g^{ij}E_{ij}$, $E_{ij}$ correspond to
equations for metric fields $\chi,\xi^{i},f$ and $\partial_{v}h_{ij}$ respectively as shown in Section \ref{sub:eom_BS}.

It is important to decouple the equations \prettyref{eq:eom_hij} for metric fields $h_{ij}$ for numerical evolutions. We find it useful to take the following redefinitions
\footnote{The $-\frac{1}{2}f\partial_{z}B$ term is not necessary for decoupling
these two equations; however, it can simplify equations in some cases.}
\begin{eqnarray}
\Pi_{B} & = & \left[\partial_{v}B-\frac{1}{2}f\partial_{z}B+\xi^{i}\partial_{i}B\right]\frac{\cosh C}{z},\label{eq:def_PI_B}\\
\Pi_{C} & = & \left[\partial_{v}C-\frac{1}{2}f\partial_{z}B+\xi^{i}\partial_{i}C\right]\frac{1}{z},\label{eq:def_PI_C}
\end{eqnarray}
which transform equations \prettyref{eq:eom_hij} into
\begin{equation}
\begin{pmatrix}\partial_{z} & \partial_{z}B\sinh C\\
-\partial_{z}B\sinh C & \partial_{z}
\end{pmatrix}\begin{pmatrix}\Pi_{B}\\
\Pi_{C}
\end{pmatrix}=\begin{pmatrix}S_{\Pi_{B}}\\
S_{\Pi_{C}}
\end{pmatrix}.\label{eq:eom_hij_PI_BC}
\end{equation}
Then we apply a further $SO\left(2\right)$ rotation for $\Pi_{B},\Pi_{C}$,
\begin{eqnarray}
\begin{pmatrix}\widetilde{\Pi}_{B}\\
\widetilde{\Pi}_{C}
\end{pmatrix} & = & \exp\left[\begin{pmatrix}0 & K\\
-K & 0
\end{pmatrix}\right]\begin{pmatrix}\Pi_{B}\\
\Pi_{C}
\end{pmatrix},\label{eq:SO_2_rotation}
\end{eqnarray}
where $K$ is defined as
\begin{eqnarray}
K & := & \int_{0}^{z}d\hat{z}\partial_{\hat{z}}B\sinh C.\label{eq:def_K}
\end{eqnarray}
Equations \prettyref{eq:eom_hij_PI_BC}, or equivalently \prettyref{eq:eom_hij},
then can be decoupled in terms of $\widetilde{\Pi}_{B},\widetilde{\Pi}_{C}$,
\begin{equation}
\partial_{z}\begin{pmatrix}\widetilde{\Pi}_{B}\\
\widetilde{\Pi}_{C}
\end{pmatrix}=\begin{pmatrix}\cos K & \sin K\\
-\sin K & \cos K
\end{pmatrix}\begin{pmatrix}S_{\Pi_{B}}\\
S_{\Pi_{C}}
\end{pmatrix}.\label{eq:eom_PI_B_C_tilde}
\end{equation}

The Hamiltonian and momentum constraints are used to evolve boundary
fields at $z=0$ and monitor numerical errors in the bulk spacetime
which can be derived from 
\begin{align}
\mathcal{H} & =G_{\mu\nu}n^{\mu}n^{\nu}=0,\label{eq:equation_of_hamiltonian_constraint}\\
\mathcal{M}_{i} & =G_{i\mu}n^{\mu}=0,\quad i=2,3,\label{eq:equations_of_momentum_constraint}
\end{align}
where $G_{\mu\nu}=0$ is the Einstein equation \prettyref{eq:EFEs}
and 
\begin{equation}
n_{\mu}=\left(0,-\ell\left(e^{\chi}fz^{2}\right)^{-1/2},0,0\right)
\end{equation}
corresponds to the normal vector pointing outward from the time-like
AdS boundary, i.e. in direction of decreasing $z$ coordinate.

In the terminology of original papers \citep{GravitationalWavesGeneral1962,GravitationalWavesGeneral1966},
equations \prettyref{eq:eom_chi}, \prettyref{eq:eom_Pi}, \prettyref{eq:eom_f},
\prettyref{eq:eom_PI_B_C_tilde} are called \emph{main} equations
or \emph{hypersurface} equations which can be integrated within a
single null surface. Equations \prettyref{eq:equation_of_hamiltonian_constraint} and \prettyref{eq:equations_of_momentum_constraint} are called \emph{supplementary}
conditions because if they are satisfied on a $z=\mathrm{const}$ hypersurface, they remain satisfied throughout the bulk spacetime. The equation $E_{vz}$ is regarded as a \emph{trivial} equation, in the sense that it is identically satisfied provided the main equations are fulfilled.

To specify the boundary conditions, it is straightforward to find the 
near boundary series solution of the equations of motion order by order as $z\rightarrow0$,
\begin{align}
f\quad & \sim\quad1+f_{3}z^{3}+\label{eq:asymp_f}\\
\chi\quad & \sim\quad\frac{1}{8}\phi_{1}^{2}z^{2}+\frac{1}{3}\phi_{1}\phi_{2}z^{3}+\cdots\label{eq:asymp_chi}\\
\xi\quad & \sim\quad\xi_{3}z^{3}+\cdots\label{eq:asymp_xi}\\
\eta\quad & \sim\quad\eta_{3}z^{3}+\cdots\label{eq:asymp_eta}\\
B\quad & \sim\quad B_{3}z^{3}+\cdots\label{eq:asymp_B}\\
C\quad & \sim\quad C_{3}z^{3}+\cdots\label{eq:asymp_C}\\
\phi\quad & \sim\quad\phi_{1}z+\phi_{2}z^{2}+\cdots\label{eq:asymp_phi}
\end{align}
where no logarithmic terms appear in the near boundary solutions.

Finally, our integration scheme for solving equations \prettyref{eq:eom_chi},
\prettyref{eq:eom_Pi}, \prettyref{eq:eom_for_xi}, \prettyref{eq:eom_PI_B_C_tilde}
and \prettyref{eq:KG-Equation} then can be described as the following
steps:
\begin{enumerate}
\item Provide fields $B,C,\phi$ and boundary values $f_{3},\xi_{3}^{i}$
at initial time slice $v=v_{0}$, and specify $\phi_{1}$, $\partial_{v}\phi_{1}$;
\item With $B,C,\phi$ known, solve equation \prettyref{eq:eom_chi} for
$\chi$;
\item With $B,C,\phi,\chi$ known, solve equations \prettyref{eq:eom_Pi}
for $P_{i}$;
\item With $B,C,\chi,P_{i}$ known, solve equations \prettyref{eq:eom_for_xi}
for $\xi^{i}$;
\item With $B,C,\phi,\chi,P_{i},\xi^{i}$ known, solve equation \prettyref{eq:eom_f}
for $f$;
\item With $B,C,\phi,\chi,P_{i},\xi^{i},f$ known, solve equation \prettyref{eq:KG-Equation}
for $\partial_{v}\phi$;
\item With $B,C$ known, integrate to get $K$ from the definition \prettyref{eq:def_K};
\item With $B,C,\phi,\chi,P_{i},\xi^{i},f,K$ known, solve equations \prettyref{eq:eom_PI_B_C_tilde}
for $\widetilde{\Pi}_{B}$, $\widetilde{\Pi}_{C}$;
\item Transform $\widetilde{\Pi}_{B}$ and $\widetilde{\Pi}_{C}$ back to
$\Pi_{B}$ and $\Pi_{C}$ by equations \prettyref{eq:SO_2_rotation};
\item Transform $\Pi_{B}$ and $\Pi_{C}$ back to $\partial_{v}B$ and $\partial_{v}C$
by equations \prettyref{eq:eom_hij_PI_BC};
\item With $B,C,f_{3},\xi_{3}^{i},\phi_{1},\partial_{v}\phi_{1}$ known,
obtain $\partial_{v}f_{3},\partial_{v}\xi_{3}^{i}$ from equations
\prettyref{eq:d1v_f3}, \prettyref{eq:d1v_xi3} and \prettyref{eq:d1v_eta3};
\item With $\partial_{v}B,\partial_{v}C,\partial_{v}f_{3},\partial_{v}\xi_{3}^{i}$
all known, integrate them to next time slice $v=v_{0}+\delta v$ by
fourth-order Runge-Kutta method;
\item Repeat the above process.
\end{enumerate}
The boundary conditions are all imposed on the conformal boundary
as listed in \prettyref{eq:asymp_f}, \prettyref{eq:asymp_chi}, \prettyref{eq:asymp_xi},
\prettyref{eq:asymp_eta}, \prettyref{eq:asymp_B}, \prettyref{eq:asymp_C}
and \prettyref{eq:asymp_phi}. We use spectral methods \citep{SpectralMethodsMATLAB2000,ChebyshevFourierSpectral2001,SpectralMethodsEvolution2007,SpectralMethodsTimedependent2007,SpectralMethodsNumerical2009,SpectralMethodsAlgorithms2011,SpectralMethodsFundamentals2011}, typically using 22 Chebyshev polynomials in the radial direction $z$ and 330 or 512 Fourier modes
on each spatial $x^{i}$ direction. The time evolution is performed with a fixed time-step $\delta v = 1/100$ using fourth-order Runge-Kutta scheme. For the finest resolution considered, the total computational cost was less than five days of running.

\subsection{Locate the Apparent Horizon\label{subsec:Locate-the-Apparent}}

We display our equations for deriving the location of the apparent horizon
and some numerical techniques here. The apparent horizon is located
at where the expansion of the congruence of outgoing null geodesics
vanishes \citep{RelativistsToolkitMathematics2004}. Suppose the null
congruence is generated by the normal vector
\begin{equation}
k_{a}=\mu\nabla_{a}\Phi
\end{equation}
for some scalar $\Phi\left(x^{\mu}\right)$ and $\mu\left(x^{\mu}\right)$.
Demanding that the expansion vanishes on a hypersurface makes it an
apparent horizon. The null condition of the normal vector $k_{a}k^{a}=0$
gives the time derivative of $\Phi$,
\begin{equation}
\partial_{v}\Phi=\frac{1}{2}f\partial_{z}\Phi-\xi^{i}\partial_{i}\Phi+\frac{1}{2}\Theta^{ij}\partial_{i}\Phi\partial_{j}\Phi\left(\partial_{z}\Phi\right)^{-1}.\label{eq:dt_Phi_AH}
\end{equation}

Requiring the null geodesics to be affinely parameterized gives the
condition $\nabla^{a}\Phi\nabla_{a}\mu=0$. Substituting these two equations
into the vanishing expansion equation $\theta=\nabla_{a}k^{a}=0$
and requiring it to be satisfied at a hypersurface $\Phi:=z-H\left(v,x,y\right)=0$ leads to the equation which determines the location of the apparent
horizon 
\begin{widetext}
\begin{equation}
0=-\partial_{H}\left[\Theta^{ij}\right]\partial_{i}H\partial_{j}H+\frac{d-1}{2H}\left[-f+\Theta^{ij}\partial_{i}H\partial_{j}H\right]+\partial_{i}\left[\Theta^{ij}\partial_{j}H\right]-\partial_{i}\xi^{i}+\partial_{H}\xi^{i}\partial_{i}H.\label{eq:equation_of_apparent_horizon}
\end{equation}

In our case, we take $d=3.$ Since this apparent horizon equation
\prettyref{eq:equation_of_apparent_horizon} is generally a nonlinear
elliptic equation in a curved space, we solve it by the Newton-Raphson
iteration method. First, we need to linearize equation \prettyref{eq:equation_of_apparent_horizon}
and solve the resulting linear system of equations
\begin{align}
\mathcal{J}\delta H & =-\mathcal{E}\label{eq:AH_Newton_Raphson_Linear_EQs}
\end{align}
to get a correction $\delta H$, where $\mathcal{J}$ is the Jacobian
and $\mathcal{E}$ is essentially equation \prettyref{eq:equation_of_apparent_horizon}.
Second, update $H_{n+1}=H_{n}+\delta H$ and iterate these steps again
until $H$ converges to the desired result. The explicit form for
$\mathcal{J}$ and $\mathcal{E}$ are given by
\begin{eqnarray}
\mathcal{J} & = & \Theta^{ij}\partial_{i}\partial_{j}+2\left[\Theta^{ij}H^{-1}-\partial_{H}\Theta^{ij}\right]\partial_{i}H+\left[\partial_{i}\Theta^{ij}+\partial_{H}\xi^{j}\right]\partial_{j}\nonumber \\
 &  & +\left(H^{-1}\partial_{H}\Theta^{ij}-\Theta^{ij}H^{-2}-\partial_{H}^{2}\Theta^{ij}\right)\left(\partial_{i}H\partial_{j}H\right)+\left[\partial_{H}\partial_{i}\Theta^{ij}+\partial_{H}^{2}\xi^{j}\right]\partial_{j}H\nonumber \\
 &  & +\partial_{H}\Theta^{ij}\left(\partial_{i}\partial_{j}H\right)-\left(\partial_{H}\partial_{i}\xi^{i}\right)+fH^{-2}-H^{-1}\partial_{H}f\\
\mathcal{E} & = & \Theta^{ij}\partial_{i}\partial_{j}H+\left[H^{-1}\Theta^{ij}-\partial_{H}\Theta^{ij}\right]\partial_{i}H\partial_{j}H+\left[\partial_{i}\Theta^{ij}+\partial_{H}\xi^{j}\right]\partial_{j}H-\partial_{i}\xi^{i}-fH^{-1}.
\end{eqnarray}
\end{widetext}

To numerically solve this equation \prettyref{eq:equation_of_apparent_horizon},
we implement spectral methods which take advantage of global
data to interpolate functions and their derivatives. This often leads
to a non-sparse matrix and its dimensions grow very quickly when discretizing
$\mathcal{J}$ in two or higher dimensions. Thus a direct linear solver,
like taking a direct inversion or LU decomposition, is inefficient. A multi-domain decomposition method \citep{MultidomainSpectralMethod2003},
or a Schur complement domain decomposition \citep{IterativeMethodsSparse2003,NewPseudospectralCode2022}
with direct solver methods are both viable options. Nevertheless,
such domain decompositions impose distinct boundary conditions across
each subdomain, thereby increasing the overall complexity. We instead
use the iteration method \citep{IterativeMethodsSparse2003} with
a simple finite difference preconditioner \citep{SpectralMethodsProblems1980,SpectralMethodsFundamentals2011}
in only a single domain and implement these in a matrix-free way.
We find the \emph{BICGSTAB} or \emph{GMRES} method as one of the Krylov
subspace iteration methods with a second-order finite difference preconditioner
surprisingly fast. See also a similar treatment in a recent paper \citep{TailsBulkGravitational2025}.

To be specific, we need to find a suitable matrix $\mathcal{M}$ as a
good approximation of $\mathcal{J}$ and substitute it into equation
\prettyref{eq:AH_Newton_Raphson_Linear_EQs} to obtain left preconditioning
\begin{equation}
\left(\mathcal{M}^{-1}\mathcal{J}\right)\delta H=-\left(\mathcal{M}^{-1}\mathcal{E}\right),
\end{equation}
or right preconditioning 
\begin{equation}
\left(\mathcal{J}\mathcal{M}^{-1}\right)\left(\mathcal{M}\delta H\right)=-\mathcal{E}.
\end{equation}
It is ideal for the operators $\mathcal{M}^{-1}\mathcal{J}$ or $\mathcal{J}\mathcal{M}^{-1}$
to be close to the identity so that the iteration may converge rapidly.
To construct the preconditioner $\mathcal{M}$, the derivative operators
in $\mathcal{J}$ are replaced by the corresponding finite-difference
operators, which are obtained from the following relations,
\begin{align}
\left(\partial_{x}^{2}f\right)\left(x_{j}\right) & =\frac{f_{j+1}-2f_{j}+f_{j-1}}{2\Delta x^{2}},\\
\left(\partial_{x}f\right)\left(x_{j}\right) & =\frac{f_{j+1}-f_{j-1}}{2\Delta x}.
\end{align}
for a given function $f$ at a discretized grid $x_{j}.$

Finally, we interpolate the metric fields along $z$ direction through
the accurate and efficient Barycentric interpolation method \citep{BarycentricLagrangeInterpolation2004}
whenever Newton-Raphson iteration reaches the field values outside the
Gauss-Chebyshev-Lobatto collocation points.

\bibliography{refs,xampl}
\end{document}